\documentclass[letterpaper,twocolumn,10pt]{article}
\usepackage{usenix}

\usepackage[dvipsnames,table]{xcolor}
\usepackage{threeparttable}
\usepackage{tcolorbox} 
\usepackage{tabularx}   % For flexible-width columns
\usepackage{graphicx}  
\usepackage{makecell}  
\usepackage{amssymb}  
\usepackage{booktabs}
\usepackage{diagbox}
\usepackage{array}
\usepackage{multirow}
\usepackage{ifthen}
\usepackage{xurl}
\usepackage{amsmath,amssymb,amsfonts}
\usepackage{float} 
\usepackage{enumitem}
\usepackage{tikz}
\usepackage{ragged2e}
\microtypesetup{spacing=false}
\usepackage{pifont}
\usepackage{wasysym}
\usepackage{fontawesome5}
\usepackage{hyperref}
\newcolumntype{L}[1]{>{\RaggedRight\arraybackslash}m{#1}}
\newcolumntype{C}[1]{>{\centering\arraybackslash}m{#1}}
\newcolumntype{P}[1]{>{\RaggedRight\arraybackslash\hspace{0pt}}p{#1}}

\newcommand{\diffbadge}[3]{%
\tikz[baseline=(n.base)]{
\node[
rounded corners=1.3pt,
fill=#2,
text=#3,
inner xsep=3.0pt,
inner ysep=0.9pt,
font=\scriptsize\bfseries
] (n) {#1};}
}

\newcommand{\dA}{\diffbadge{1}{teal!35}{teal!0!black}}
\newcommand{\dB}{\diffbadge{2}{green!45}{green!0!black}}
\newcommand{\dC}{\diffbadge{3}{yellow!55}{brown!0!black}}
\newcommand{\dD}{\diffbadge{4}{orange!50}{orange!0!black}}
\newcommand{\dE}{\diffbadge{5}{red!45}{red!0!black}}

\newcommand{\lowc}{\Circle}        % empty circle
\newcommand{\medc}{\LEFTcircle}    % half circle
\newcommand{\highc}{\CIRCLE}       % filled circle

\newcommand{\halfcheck}{%
  \mathrel{%
    \ooalign{%
      $\checkmark$\cr
      \hfil\raise0.2ex\hbox{$\backslash$}\hfil\cr
    }%
  }%
}

\begin{document}
\newboolean{showcomments}
\setboolean{showcomments}{true}
\ifthenelse{\boolean{showcomments}}
{\newcommand{\mynote}[2]{
  \fbox{\bfseries\sffamily\scriptsize#1}
  {\small
  $\blacktriangleright$
  \textsf{{{\em #2}\bf}}
  $\blacktriangleleft$}}
}

\newcommand{\van}[1]{{\color{blue}{#1}}}
\newcommand{\fc}[1]{\mynote{fc}{\textcolor{BrickRed}{#1}}}
\newcommand{\crud}[1]{\mynote{CR}{\textcolor{Green}{#1}}}
\newcommand{\tina}[1]{\mynote{tina}{\textcolor{red}{#1}}}

%don't want date printed
% \date{}

\title{\Large \bf SoK: Systematizing Generation, Characteristics, and Defenses\\ in LLM-Generated Phishing}

%for single author (just remove % characters)
\author{
{\rm Fengchao Chen}\\
Monash University\\
    CSIRO Technology
\and
{\rm Tingmin Wu}\\
CSIRO Technology
\and
{\rm Van Nguyen }\\
Monash University
\and
{\rm Carsten Rudolph}\\
Monash University
% copy the following lines to add more authors
% \and
% {\rm Name}\\
%Name Institution
} % end author

\maketitle

%-------------------------------------------------------------------------------
\begin{abstract}
The rapid advancement of Large Language Models (LLMs), with their growing abuse in phishing, has enabled phishing content, including deceptive pretexts and other persuasive elements, to be generated at a scale difficult to achieve manually. This growing misuse of LLMs in phishing raises questions about potential LLM-driven changes in phishing characteristics, associated security implications for users, and the resulting challenges to existing defenses. While prior research has examined individual dimensions, including threats of LLM abuse, LLM-driven phishing, user susceptibility to phishing, and phishing defenses, these efforts remain fragmented and have not been consolidated into a comprehensive understanding of how these dimensions interrelate. To address this research gap, we provide a systematic examination of the LLM-generated phishing landscape, including a taxonomy of LLM manipulation methods, characteristics associated with LLM-generated phishing threats, and a taxonomy of defenses aligned with these manipulation methods. We also benchmark five academic and five industrial phishing detectors across datasets associated with different LLM-based generation methods. We further extract insights and research gaps that suggest promising directions for future research in this growing area. Our findings underscore the need for countermeasures that are evaluated across LLM generators and manipulation strategies.
Our work provides a systematic foundation for studying LLM-generated phishing, enabling more consistent comparison and evaluation across the community.

\end{abstract}

\begin{figure*}[ht]
    \centering
\includegraphics[width=0.95\linewidth]{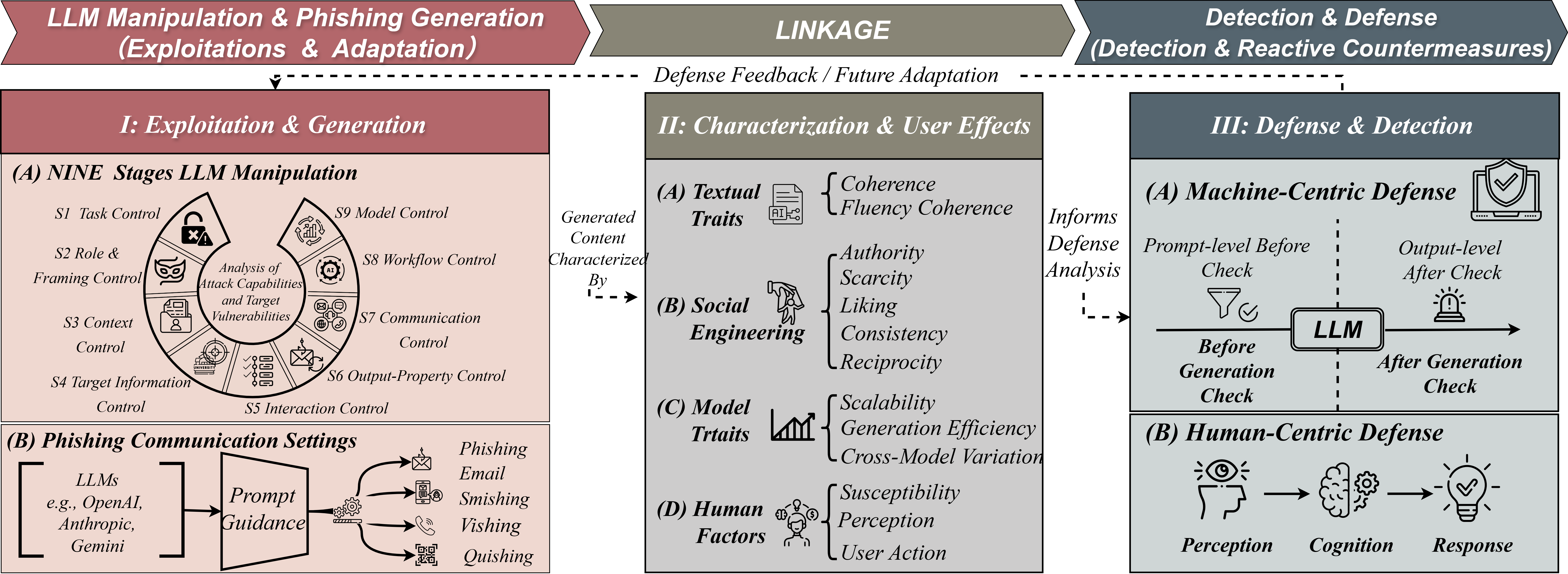}
    \caption{Overview of our systematization of LLM-generated phishing. (\S\ref{RQ1}) organizes phishing-generation methods by the LLM-related factor under attacker control; (\S\ref{RQ2}) characterizes the resulting phishing content and reported user-side effects; and (\S\ref{RQ3}) systematizes defenses and evaluates detector robustness across heterogeneous LLM-generated phishing settings. }
    % The (\S\ref{RQ1}) stages denote manipulation loci and do not represent a required attack sequence. % Dataset availability and evaluation metrics are examined throughout the lifecycle (RQ4).
    \label{fig:GED}
\end{figure*}
%-------------------------------------------------------------------------------
\section{Introduction}\label{introduction}
``Your Paypall account has been suspended; please verify it within 30 minutes.'' This seemingly routine message illustrates a phishing lure through which adversaries use persuasive communication to induce harmful actions (e.g., downloading malware)~\cite{bethany2024deciphering,carroll2022good}. Phishing content commonly consists of a deceptive pretext that establishes a fabricated scenario or identity (e.g., the fabricated account suspension and claimed identity ``Paypall''), behavioral inducements that specify the action (e.g., ``verify your account''), persuasive cues that motivate compliance~(e.g., ``within 30 minutes''), and actionable artifacts that enable the action (e.g., URLs)~\cite{fette2007learning,stojnic2021phishing}.

While actionable artifacts provide the means to perform the induced action, the surrounding textual content remains the primary medium for many phishing communications~\cite{carroll2022good}. More concerningly, the proliferation of Large Language Models (LLMs) further expands the capacity for generating textual phishing content, enabling rapid production of diverse and contextually tailored content~\cite{99_global_phishing}. 
Industry reports indicate broader AI adoption in phishing: 82.6\% of analyzed phishing emails showed signs of AI use~\cite{Knowbe4}, while AI-enabled phishing operations achieved click-through rates of up to 54\%, compared with roughly 12\% for traditional campaigns~\cite{microsoft_phishing_report}. 
%As reported, over 80\% of analyzed phishing emails involved LLM-assisted rephrasing~\cite{Knowbe4}, where the rephrased textual content is more credible to users, increasing the likelihood of user engagement and yielding attack effectiveness gains of up to 450\% over human-written counterparts~\cite{microsoft_phishing_report}. 
The real-world impact is already evident, with phishing involving LLMs associated with over \$10.3 million in reported losses in 2025~\cite{FBI_IC3_2025}. These observations motivate an understanding of how LLM-generated phishing differs from human-written phishing, and what changes account for different attack effects.

Existing surveys on malicious uses of LLMs examine broad misuse cases, attack targets and affected systems, and LLM-enabled activities across different forms of cybercrime~\cite{shanmugarasa2025sok,schroer2025sok}, but provide limited analysis of how LLM misuse manifests within specific threat domains such as phishing. Prior studies on human factors primarily emphasize user-side susceptibility, such as limited security training and cognitive factors~\cite{desolda2021human,jabir2025phishing,spicer2026makes,zhuo2023sok}, while offering limited insights on what phishing characteristics introduced by LLMs affect user decision-making and actions. Defense-oriented works have reviewed phishing datasets across phishing modalities (e.g., webpages, emails)~\cite{das2019sok}, defenses against non-LLM-generated phishing~\cite{das2019sok,dou2017systematization}, and LLMs for phishing security~\cite{sivaneswaran2026systematic}, but provide limited systematization of defenses specifically targeting LLM-generated phishing attacks. The closest recent SoK, PHILTER~\cite{alamsok}, uses LLM-assisted reasoning to evaluate whether AI-based phishing-website detectors meet real-world functional and security requirements, focusing on LLMs for phishing security rather than LLM-generated phishing defenses. Despite these advances, existing works remain fragmented across phishing research, with limited connections across different areas of phishing research, leaving the emerging threat landscape and security implications of LLM-generated phishing insufficiently understood.

To address this gap, we conduct a systematization of knowledge on the phishing landscape of LLMs. We approach LLM-generated phishing from a holistic perspective by organizing the LLM abuse methods for phishing content generation, systematizing the resulting phishing characteristics and corresponding effects on users, and assessing the challenges that LLM-induced changes pose to existing phishing defenses. The analysis first formalizes LLM-exploiting practices for phishing generation into nine non-exclusive stages defined by the LLM-related factor under attacker control, spanning input, interaction, output, workflow, and model-level control (\S\ref{RQ1}).
We then systematize the resulting threats through a taxonomy spanning textual and social engineering characteristics, model-level traits, and their effects on user perception and behavior~(\S\ref{RQ2}). Finally, we organize existing machine- and human-centered defenses, evaluate the robustness of detectors from both academia and industry across different LLM-generated phishing content, generation strategies, and phishing settings~(\S\ref{RQ3}). 
Furthermore, we identify key insights and research gaps, and outline the open challenges for future phishing defenses.
We refer to Fig.~\ref{fig:GED} for an overview of the key dimensions and properties of LLM-generated phishing content. In summary, we make the following \textbf{contributions}:

\begin{itemize}[leftmargin=*]
    
    % \item We perform the first comprehensive literature review on LLM-driven phishing campaigns, focusing on Email and SMS phishing, pretext-based QR code phishing (Quishing), scripts for voice phishing (Vishing), and poisoned content for image phishing (for text-to-image models), with a taxonomy that highlights the asymmetric trajectories of offense and defense mechanisms.
    % \item We outline nine stages that describe the methods adversaries leverage to explore the ethical threshold of LLM compliance, uncovering the evolving trajectories and potential escalation paths of LLM-generated phishing and related attack mechanisms.
    % \item We contribute comparative insights that demonstrate how LLM-generated phishing achieves quantitative effectiveness in phishing campaigns and highlight an asymmetry in defending against it.
    % \item We highlight key insights, challenges, and priority research directions specific to different stages of LLM-generated phishing.
    \item \textbf{Nine-Stage LLM Manipulation Framework:} %We introduce a nine-stage framework for systematizing LLM manipulation methods in phishing content generation, facilitating consistent characterization and comparison across attack methodologies. 
    We introduce a nine-stage framework for systematizing LLM manipulation methods by the primary factor under attacker control; the stages are composable and do not represent a required attack sequence or capability ranking.

    \item \textbf{Characterization of LLM-Generated Phishing:} We provide a systematic characterization of LLM-generated phishing across generation strategies, phishing communication settings, and LLM generators, facilitating deeper understanding of variations in phishing characteristics, comparative analysis, and the development of more robust defenses.
    % and consolidating reported user-side effects from prior studies.

    \item \textbf{Defense Landscape and Benchmarking:} We provide a generation-aware framework for organizing and evaluating existing phishing defenses, supporting systematic comparison of defense coverage and robustness against heterogeneous LLM-generated phishing.
\end{itemize}

% \textbf{Key insights} from our work include:
% \begin{itemize}[leftmargin=*]
%     \item LLM-TCP generation methods show a trend that evades models' change to attack users' cognition  
%     \item key insights on LLM-TCP attack properties
%     \item the development of defense has distinctive differences with the advancement of LLM-TCP generation
% \end{itemize}

% \noindent \textit{Organization}. Section~\ref{methods_of_paper_selection} outlines our methodology and research questions; We present our taxonomy in Section~\ref{RQ1}, mapping the evolution of LLM-generated phishing. Then, characteristics distinguishing attack patterns are shown in Section~\ref{RQ2} and a categorization of proposed defenses in Section~\ref{RQ3}. 
% Section~\ref{data_metric} provides context on datasets and evaluation metrics used in each stage of LLM-generated phishing. Sections~\ref{further_directions} and Section~\ref{conclusion} discuss future research directions and conclusions. Insights and gaps throughout the paper highlight our findings.

To the best of our knowledge, this is the first work to systematically analyze the interdependent components of the LLM-generated phishing landscape. Our systematization provides a foundational framework for understanding the emerging threat of LLM-generated phishing and associated implications for existing defenses, guiding future research toward building robust phishing detection and defense mechanisms.
% Our systematization provides a foundational framework for understanding how LLMs scale phishing attacks, how different phishing characteristics are from existing phishing, and how existing defense assumptions and coverage are challenged under changes, then guiding future research toward building robust phishing detectors.
% This work takes a bottom-up approach to make it accessible to general readers interested in this area. 
We release the list of work, datasets, and code in the GitHub repository~\cite{sok_git}.

%-------------------------------------------------------------------------------
\section{Overview}\label{methods_of_paper_selection}
In this section, we define our research scope, paper filtering approach, and discuss the key differences from existing surveys on phishing in the age of LLMs.

\noindent \textbf{1) Research Scope.} In this work, we focus on \textit{LLM-generated textual content} that functions as phishing communication or facilitates a phishing interaction. This includes deceptive pretexts, behavioral inducements, persuasive cues, and textual actionable artifacts such as URL strings. Here, \textit{LLM-generated phishing content} refers specifically to such language-based content, rather than the broader set of components involved in a phishing attack. We consider an attacker who can access an LLM through prompting, workflow orchestration, or model/data modification to generate or transform phishing content. We focus on the generation, characteristics, and content-focused defenses of LLM-generated phishing, excluding broader phishing attack operations such as domain registration, message delivery, and credential harvesting.
% % The threat surfaces are the LLM-related factors represented by \textit{S1--S9}; these stages assume different access capabilities and need not all be available to a single attacker. A pretext is therefore one component of phishing content, rather than a synonym for all content generated or used in the attack. 

We consider different phishing communication settings in which LLM-generated content can be used, including phishing emails, SMS-based phishing (Smishing), and scripts or messages in voice phishing (Vishing). We also include textual artifacts used in QR-code phishing (Quishing), such as URL strings encoded in QR codes, but exclude the generation of QR-code images themselves. 
% Generating a URL string does not include registering the domain, hosting the destination, or operating the associated infrastructure. 
We likewise exclude phishing content whose effectiveness substantially depends on visual, structural, and document-level factors that go beyond the linguistic content considered here, including webpages, PDFs, attachments, or images. 
\begin{table}[!t]
\caption{Overview of Studies on LLM-Generated Phishing.}
\label{tab:overview_all_studies}
\centering
\tiny
\setlength{\tabcolsep}{0.8pt}
\renewcommand{\arraystretch}{0.75}
\resizebox{\columnwidth}{!}{
\begin{tabular}{p{0.5cm} p{1.7cm} p{0.8cm} p{2.3cm} p{2cm} p{0.5cm}}
\toprule
\textbf{Year} & \textbf{Venue} & \textbf{Cat.} & \textbf{Focus} & \textbf{Modality} & \textbf{Work} \\
\midrule

2021 & IET Conf. Proc. & G & Adversarial generation & Spear-Phishing Email & \cite{khan2021offensive} \\
\cmidrule{1-6}

2022 & BWCCA & G & Personalized generation & Phishing Email & \cite{guo2022generating} \\
2022 & arXiv & G & Targeted generation & Phishing Email & \cite{karanjai2022targeted} \\
\cmidrule{1-6}

% 2023 & arXiv & G & Personalized spear-phishing & Spear-Phishing Email & \cite{hazell2023large} \\
2023 & IWSPA@CODASPY & G & Adversarial robustness & Phishing Email & \cite{mehdi2023adversarial} \\
2023 & WithSecure Intell. & G & Malicious prompt engineering & Phishing Email & \cite{patel2023creatively} \\
2023 & IEEE BigData & G & Persuasion prompting & Conversational phishing  & \cite{singh2023exploiting} \\
2023 & SmartCity/DependSys & G & Phishing rephrasing & Phishing Email & \cite{utaliyeva2023chatgpt} \\
2023 & IJCIC & A & Victim simulation & Conversational phishing  & \cite{asfour2023harnessing} \\
2023 & EuroS\&PW & A & Cognitive-bias study & Phishing Email & \cite{sharma2023well} \\
2023 & IEEE Access & G+A+D & Phishing generation & Phishing Email & \cite{heiding2024devising} \\
\cmidrule{1-6}
2024 & Engineering Proc. & G & Voice phishing script & Vishing & \cite{toapanta2024ai} \\
2024 & IEEE BigData & G & Phishing generation & Phishing Email & \cite{fairbanks2024generating} \\
2024 & ACM TALLIP & G & Phishing generation & Phishing Email & \cite{guo2024x} \\
2024 & ICLR & G & Neural phishing & Conversational phishing & \cite{panda2024teach} \\
2024 & ISDFS & G & Prompt-injection generation & Smishing & \cite{shibli2024abusegpt} \\
2024 & arXiv & G & Persuasion-based augmentation & Smishing & \cite{shim2024persuasion} \\
2024 & ITASEC & G & Phishing generation & Phishing Email & \cite{greco2024david} \\
2024 & MCNA & A & Social-engineering analysis & Phishing Email & \cite{alahmed2024exploring} \\
2024 & arXiv & A & AI vs human spear-phishing & Smishing & \cite{francia2024assessing} \\
2024 & Artif. Intell. Rev. & A & Capability review & Phishing Email & \cite{schmitt2024digital} \\
2024 & SSRN & D & Detection in Healthcare & Phishing Email & \cite{elongha2024detecting} \\
2024 & ARES & D & LM-vs-human detection & Phishing Email & \cite{gryka2024detection} \\
2024 & BDAI & D & Multi-source phishing detection & Phishing Email/Smishing & \cite{mahendru2024securenet} \\
2024 & arXiv & D & LLM-reasoned indicators & Spear-Phishing Email & \cite{nahmias2024prompted} \\
2024 & Research Square & D & Cybersecurity policy drafting & Spear-Phishing Email & \cite{quinn2024applying} \\
2024 & Computers & D & AI-text abuse detection & Phishing Review & \cite{wani2024ai} \\
2024 & arXiv & G+A & Phishing evolution & Phishing Email & \cite{chen2024adapting} \\
2024 & ECAI & G+A & AI-assisted prompting & Phishing Email & \cite{emanuela2024ai} \\
2024 & IEEE BigData & G+A & Stealth rewriting/detector eval. & Phishing Email & \cite{afane2024next} \\
2024 & S\&P & G+D & phishing generation & Phishing Email & \cite{roy2024chatbots} \\
2024 & Electronics & G+A+D & indicator analysis/detector eval. & Phishing Email & \cite{eze2024analysis} \\
\cmidrule{1-6}
2025 & Information Fusion & G & Critique-guided refinement & Spear-Phishing Email & \cite{qi2025spearbot} \\
2025 & AsiaCCS & G & Quishing\&LLM phishing & Quishing + Phishing Email & \cite{weinz2025impact} \\
2025 & AsiaCCS & G & Voice phishing generation & Vishing & \cite{figueiredo2025sounds} \\
% 2025 & IEEE Netw. Lett. & G & Quishing exemplification & Quishing & \cite{akram2025exemplifying} \\
2025 & AAAI & G & AR-based social engineering & Conversational phishing  & \cite{bi2025feasibility} \\
2025 & EICC & G & Malicious-prompt robustness & Phishing Email/Website & \cite{ccetin2025exploring} \\
2025 & USENIX & G & Retrieve augmented phishing & Spear-phishing Email & \cite{kim2025llms} \\
2025 & arXiv & G & Semantic obfuscation & Vishing & \cite{li2025talking} \\
2025 & AISec & G & Synthetic benchmark & Phishing Email & \cite{pajola2025phishgen} \\
2025 & IMC & A & In-the-wild traits & BEC Phishing Email & \cite{hao2025spammers} \\
2025 & SIGMIS-CPR & A & User susceptibility & Phishing Email & \cite{olea2025evaluating} \\
2025 & CHI & A & User susceptibility & Conversational Phishing & \cite{veisi2025user} \\
2025 & Electronics & D & Machine learning detector & Phishing Email & \cite{brissett2025machine} \\
2025 & NDSS & D & Trigger-tag defense & Phishing Email & \cite{pang2025paladin} \\
2025 & ESWA & D & Stylometric detector eval. & Phishing Email & \cite{opara2025evaluating} \\
2025 & HCII & D & Behavioral defense & Phishing Email & \cite{malloy2025training} \\
2025 & arXiv & G+D & Knowledge-grounded detector & Spear-Phishing Email & \cite{liu2025pimref} \\
2025 & arXiv & G+D & Multi-agent detector & Phishing Email+URL+Head & \cite{xue2025multiphishguard} \\
2025 & IEEE Access & G+D & Lateral-phishing detector & Phishing Email & \cite{bethany2025lateral} \\
\cmidrule{1-6}
2026 & USENIX & G & Personalized phishing & Spear-Phishing Email & \cite{czybik2026large} \\
2026 & ACL & G & Multi-turn phishing trajectories & Conversational phishing & \cite{tewari2026trust} \\
2026 & arXiv & G & Synthetic benchmark & Phishing Email & \cite{toth2026phishspamvalidgenerating} \\
2026 & $^{*}$CHB & A & Cognitive Personalization & Phishing Email & \cite{malloy2025improving} \\
2026 & LREC & A & Intent characterization & Spear-Phishing Email & \cite{popovic2026dideco} \\
2026 & Electronics & D & Federated detector & Phishing Logs & \cite{malik2026qfedphish} \\
2026 & arXiv & G+A+D & Context-aware personalization & Spear-Phishing Email & \cite{vafa2026context} \\

\bottomrule
\end{tabular}
}
\begin{tablenotes}[flushleft]
\scriptsize
\item \textbf{Category (Cat.)}: Generation (G), Analysis (A), and Defense (D). Combined labels such as G+A, A+D, G+D, and G+A+D indicate that the same study is included in multiple parts of this SoK. \textbf{$^{*}$CHB}: Computers in Human Behavior.
\end{tablenotes}
\end{table}

\noindent \textbf{2) Paper screening process.} The initial search was executed via generic citation databases Scopus and Google Scholar, as well as databases focused on computer science (e.g., ACM Digital Library, IEEE Xplore). The focus was on research published in the top conferences (e.g., \textit{IEEE S\&P, ACM CCS, USENIX Security, NDSS, CHI}) and journals (e.g., \textit{TDSC, TIFS, Computer \& Security}). In light of practical considerations, we also included discussions on representative articles published recently on \textit{ArXiv} to ensure breadth and timeliness. 

For a focused but comprehensive search for significant work, the scope has been confined to \textit{``phishing content''} first to generally retrieve the main works in this area. \textit{``fake content''} is excluded to avoid substantial noise, such as Deepfake~\cite{rossler2019faceforensics}. Related retrieval terms such as \textit{``spam'', ``malicious'', and ``fraudulent''} are included to broaden coverage of phishing-related work. Similarly, \textit{``text''} and \textit{``pretext''} are used as content-related retrieval terms; they are not treated as semantic synonyms for \textit{``content''}. To retrieve LLM-authorship studies, we use keywords \textit{``large language models'', ``AI-driven'', ``Generative AI (GAI) generated'', and ``synthetic''}. The abbreviation \textit{LLMs} resulted in ambiguous search results and was therefore excluded. Finally, we iteratively used 48 search phrases, such as ``AI-driven phishing content'', to retrieve articles.

The initial search returned 10,400 records. We excluded pre-2018 works (n=3,800, before modern LLMs like GPT) and removed 4,624 duplicates using Zotero and manual verification, yielding 1,976 unique papers. Two authors independently screened the titles and abstracts. Disagreements were resolved through discussion (\textit{Cohen's $\kappa = 0.73$}). We excluded work on generic phishing, LLM-based detection without LLM-generated phishing, and papers containing only tangential mentions. Screening retained 105 papers; snowballing added 6, yielding 111 for full-text assessment. We included studies only if they: (i) directly used LLMs for phishing generation (e.g., phishing emails, URL strings in Quishing); (ii) provided methodological substance (e.g., datasets, prompts, metrics); and (iii) were technical contributions rather than editorials or tutorials. The resulting 56 core papers support our taxonomy, characterization, defense analysis, and research gaps (Table~\ref{tab:overview_all_studies}).

% (Fig.~\ref{appendix:fig_PRISMA})
\begin{table*}[!t]
\caption{Systematization of LLM-based phishing generation methods}
\label{tab:LLM_generation_rq1}
\centering
\scriptsize
\setlength{\tabcolsep}{0.8pt}
\renewcommand{\arraystretch}{0.7}
\begin{threeparttable}

\resizebox{\linewidth}{!}{
\begin{tabular}{
C{0.6cm}
L{1.5cm}|
L{1.5cm}
L{4.5cm}
L{2.7cm}
L{4.5cm}
L{2.9cm}
C{0.5cm}
C{0.6cm}
C{0.7cm}
C{0.7cm}
P{1.3cm}
}
\toprule
\multirow{2}{*}{\textbf{Stage}} &
\multicolumn{1}{L{1.5cm}}{\multirow[c]{2}{*}{\makecell[c]{Manipulation\\Category}}} &
\multicolumn{1}{L{1.5cm}}{\multirow[c]{2}{*}{\makecell[c]{Implementation\\Mode}}}&
\multicolumn{4}{c}{\textbf{Attack Methodology}} &
\multicolumn{4}{c}{\textbf{Attack Properties}} &
\multirow{2}{*}{\textbf{Works}} \\
\cmidrule(r){4-7}\cmidrule(r){8-11}
&
\multicolumn{1}{L{1.5cm}}{} &
&
\textbf{Attack Tactics} &
\textbf{Exploited AI Surface} &
\textbf{Attack Goal} &
\textbf{Target LLMs} &
\textbf{Per.} &
\textbf{Aut.} &
\textbf{Prod.} &
\textbf{Diff.} &
\\
\midrule

S1
& \multicolumn{1}{c|}{\multirow[c]{14}{*}{\makecell[c]{Request \& \\Interaction \\Control}}}
& \multicolumn{1}{c}{\multirow[c]{14}{*}{\makecell[c]{Human\\Directed}}}
& Safety Controls
& Prompt Processing
& Obtain phishing content directly
& \makecell[l]{ChatGPT, GPT-4, etc.}
& \lowc
& \lowc
& \faEnvelopeOpenText
& \dA
& \cite{eze2024analysis,ccetin2025exploring} \\

S2
&
&
& Generative-critique role prompting
& Safety Controls
& Legitimize and refine phishing assistance
& \makecell[l]{GPT-4}
& \lowc
& \lowc
& \faEnvelopeOpenText
& \dA
& \cite{qi2025spearbot} \\

S3
&
&
& Scenario- or pretext-driven prompting
& Prompt Processing
& Construct plausible lures and pretexts
& \makecell[l]{GPT-3.5/4, Qwen-2.5, etc.}
& \medc
& \lowc
& \faEnvelopeOpenText
& \dB
& \cite{emanuela2024ai,karanjai2022targeted,toth2026phishspamvalidgenerating} \\

S3
&
&
& Persuasion-conditioned prompt generation
& Prompt Processing
& Encode persuasive cues into lures
& \makecell[l]{Unspecified}
& \medc
& \lowc
& \faEnvelopeOpenText
& \dB
& \cite{heiding2024devising,shim2024persuasion} \\

S3
& 
& 
& OSINT-grounded prompting
& Prompt Processing
& Craft organization-specific lures
& \makecell[l]{ChatGPT-3.5 Turbo}
& \medc
& \medc
& \faEnvelopeOpenText
& \dC
& \cite{weinz2025impact} \\

S3
&
&
& Social-engineering scenario prompting
& Prompt Processing
& Increase realism and victim compliance
& \makecell[l]{GPT-3.5, GPT-4-Turbo, etc.}
& \medc
& \medc
& \faEnvelopeOpenText
& \dC
& \cite{greco2024david,bethany2025lateral} \\

S4
& 
& 
& Profile/Retrieval-enhanced prompting
& Retrieval Context
& Scale personalization and credibility
& \makecell[l]{GPT-4-class, etc.}
& \highc
& \medc
& \faEnvelopeOpenText
& \dB
& \cite{kim2025llms,czybik2026large} \\

S4
& 
& 
& Profile-grounded prompting
& Retrieval Context
& Personalize pretexts with target context
& \makecell[l]{GPT-4, Claude 3 Haiku, etc.}
& \highc
& \highc
& \faEnvelopeOpenText
& \dC
& \cite{vafa2026context} \\

S4
&
&
& Scene-grounded multimodal interaction
& Prompt Processing
& Ground and personalize pretexts in situ
& \makecell[l]{Claude, etc.}
& \highc
& \highc
& \faEnvelopeOpenText
& \dC
& \cite{bi2025feasibility,pajola2025phishgen,liu2025pimref} \\

S5
& 
&
& Subtasks decomposition
& Safety Controls
& Get phishing from benign subtasks
& \makecell[l]{GPT-3.5, GPT-4-Turbo, etc.}
& \lowc
& \lowc
& \faEnvelopeOpenText
& \dB
& \cite{patel2023creatively,singh2023exploiting,shibli2024abusegpt} \\
\cline{1-12}

S6
& \multicolumn{1}{c|}{\multirow[c]{6}{*}{\makecell[c]{Output\\Control}}}
& \multicolumn{1}{c}{\multirow[c]{2}{*}{\makecell[c]{Human\\Directed}}}
& Stealth rewriting or paraphrasing
& Model Outputs
& Reduce detectability while preserving intent
& \makecell[l]{GPT-3.5/4, etc.}
& \lowc
& \medc
& \faEnvelopeOpenText
& \dA
& \cite{afane2024next,utaliyeva2023chatgpt} \\

S6
&
&
& Multi-turn Stealth rewriting
& Model Outputs
& Reduce detectability while preserving intent
& \makecell[l]{GPT-4o, etc.}
& \lowc
& \medc
& \faEnvelopeOpenText
& \dB
& \cite{xue2025multiphishguard} \\

\cline{3-3}\cline{4-12}

S7
&
& \multicolumn{1}{c}{\multirow[c]{2}{*}{\makecell[c]{Workflow\\Automated}}}
& Autonomous conversational Vishing
& Planning Loops
& Extract sensitive information live
& \makecell[l]{ChatGPT + TTS, etc.}
& \highc
& \highc
& \faPhone*
& \dD
& \cite{figueiredo2025sounds,toapanta2024ai} \\

S7
&
&
& Adversarial Vishing transcript rewriting
& Model Outputs
& Evade classifiers, preserve scam intent
& \makecell[l]{GPT-4o, Gemini 2.0, etc.}
& \lowc
& \highc
& \faPhone*
& \dD
& \cite{li2025talking} \\
\cline{1-12}

S8
& \multicolumn{1}{c|}{\multirow[c]{11}{*}{\makecell[c]{Workflow \&\\Model\\Control}}}
& \multicolumn{1}{c}{\multirow[c]{6}{*}{\makecell[c]{Workflow\\Automated}}} 
& LLM-generated malicious prompt search
& Generation Pipelines
& Automate malicious prompt discovery
& \makecell[l]{ChatGPT, GPT-4, etc.}
& \lowc
& \highc
& \faEnvelopeOpenText,\faLink
& \dB
& \cite{roy2024chatbots} \\

% S8
% &
% &
% & QR-based BiTB prompting
% & Safety Controls
% & Conceal links and harvest credentials
% & \makecell[l]{Gemini}
% & \lowc
% & \medc
% & \faLink
% & \dC
% & \cite{akram2025exemplifying} \\

S8
&
&
& Outcome-driven multi-turn optimization
& Planning Loops
& Adapt interactions toward compromise
& \makecell[l]{Gemini 2.5 Flash}
& \highc
& \highc
& \faEnvelopeOpenText
& \dD
& \cite{tewari2026trust} \\

S8
&
& 
& Game-theoretic generation optimization
& Generation Pipelines
& Maximize phishing quality and payoff
& \makecell[l]{GPT-2}
& \lowc
& \highc
& \faEnvelopeOpenText
& \dE
& \cite{khan2021offensive} \\

S8
&
&
& Reflective beam-search evasion rewriting
& Generation Pipelines
& Rewrite emails into evasive variants
& \makecell[l]{GPT-3.5}
& \lowc
& \highc
& \faEnvelopeOpenText
& \dE
& \cite{fairbanks2024generating} \\

\cline{3-12}

S9
&
& \multicolumn{1}{c}{\multirow[c]{4}{*}{\makecell[c]{Model\\Adapted}}}
& Phishing-corpus/cross-lingual fine-tuning
& Models
& Internalize phishing style and patterns
& \makecell[l]{GPT-2}
& \lowc
& \highc
& \faEnvelopeOpenText
& \dD
& \cite{guo2022generating,guo2024x} \\

S9
&
& 
& Training-data poisoning for neural phishing
& Data and Model Poisoning
& Trigger targeted disclosure or misbehavior
& \makecell[l]{Unspecified}
& \lowc
& \highc
& \faEnvelopeOpenText
& \dD
& \cite{panda2024teach} \\

S9
&
&
& Adversarial example augmentation
& Models
& Stress-test detectors with adversarial emails
& \makecell[l]{GPT-2, etc.}
& \lowc
& \highc
& \faEnvelopeOpenText
& \dE
& \cite{mehdi2023adversarial} \\

S9
&
&
& Iterative adversarial sample evolution
& Models
& Evolve variants and expose blind spots
& \makecell[l]{Llama 3, etc.}
& \lowc
& \highc
& \faEnvelopeOpenText
& \dE
& \cite{chen2024adapting} \\

\bottomrule
\end{tabular}
}

\begin{tablenotes}[flushleft]
\scriptsize
\item 
\lowc, \medc, \highc~increasing level of reliance on victim-specific information. \quad
\faEnvelopeOpenText ~= Phishing Email\quad
\faPhone*~= Vishing Script\quad
\faLink~= URL Strings \quad
\dA\dB\dC\dD\dE; a higher number \\represents harder generation effort and complex execution configurations. \textbf{Exploited AI Surface} is adapted from established LLM security taxonomies (OWASP GenAI Top 10~\cite{OWASP}\\ and MITRE ATLAS~\cite{MITRE_ATLAS}).
\end{tablenotes}
\end{threeparttable}
\end{table*}
\section{Systematizing LLM Manipulation for Phishing Content Generation}\label{RQ1}
% This section systematizes LLM manipulation for phishing generation through a nine-stage framework, and reviews representative methods, attack properties, and real-world misuse.

\subsection{Systematization Methodology}
\noindent \textbf{1) Data Extraction and Coding.} We first read and analyzed each study marked with $G$ in Table~\ref{tab:overview_all_studies}. For each work, we extracted the specific LLM manipulation mechanisms. We then identified the LLM-related factor manipulated by each mechanism for later categorization. For instance, role-play assigns LLMs a specific identity (e.g., a security researcher) before generating a phishing email. We therefore record role setting as the manipulated LLM-related factor, with the mechanism altering the identity assigned to the LLM, which in turn conditions LLMs' behavior during phishing content generation. The process was informed by two LLM security frameworks: MITRE ATLAS~\cite{MITRE_ATLAS} and the OWASP GenAI Top 10~\cite{OWASP}.
% We also recorded related attack goals, target LLMs, and attack properties of each manipulation mechanism for further analysis.

\noindent \textbf{2) Taxonomy Construction and Stage Assignment.} Based on the extracted manipulation mechanisms and corresponding manipulated LLM-related factors, we grouped manipulation mechanisms that operate on the same or closely related LLM-related factors into the same stage, resulting in nine stages: \textit{S1)} Task Control, \textit{S2)} Role \& Framing Control, \textit{S3)} Context Control, \textit{S4)} Target-Information Control, \textit{S5)} Interaction Control, \textit{S6)} Output-Property Control, \textit{S7)} Communication Control, \textit{S8)} Workflow Control, and \textit{S9)} Model Control shown in Table~\ref{tab:LLM_generation_rq1}. These nine stages resulted from consolidating mechanisms that manipulate the same or closely related LLM-related factors; they cover the mechanisms identified in the current corpus but are not claimed to be exhaustive for future attacks. %The stages are ordered by where attacker intervention occurs, ranging from input-level factors to interaction, output, workflow, and the models. Stage-assignment disagreements were resolved by consensus. 
Two authors independently assigned mechanisms to stages, and disagreements were resolved through discussion and consensus. 
%The numbering reflects where attacker intervention occurs, from request-level factors through attacker--LLM interaction and victim-facing output control to workflow- and model-level control; it does not rank attack sophistication, required capability, or effectiveness.
The numbering is used only to label intervention loci---request-level factors, attacker--LLM interaction, victim-facing output, workflows, and models---and does not encode temporal order or rank attack sophistication, required capability, or effectiveness.

The stages do not represent a required sequence from \textit{S1} to \textit{S9}, and manipulation mechanisms from different stages may appear in the same method. For example, a scenario-based prompt may include a direct phishing request (\textit{S1}) and role-play (\textit{S2}), while further incorporating target-specific information, such as the recipient's individual profile or real-time context (\textit{S4}). When a study manipulates multiple factors, we code each mechanism separately; for summary tables that require a single primary stage, we use the factor that the study primarily manipulates and evaluates, not the highest-numbered stage. Accordingly, the example is assigned to \textit{S4} as its primary stage because victim-specific information is the focal manipulated factor, while \textit{S1} and \textit{S2} are treated as supporting mechanisms. Compared with other alternatives (e.g., prompt-based classification), our approach separates the underlying manipulation from the prompt form used to implement it, enabling consistent classification across different prompt designs. 
% \textit{S3} captures group-, or organization-level context, whereas \textit{S4} requires victim-specific or real-time target information. \textit{S5} concerns the attacker--LLM interaction used to elicit content across turns, whereas \textit{S7} concerns the victim-facing communication structure and \textit{S8} concerns automated orchestration across prompts, tools, feedback, or agents. \textit{S6} changes properties of generated outputs, whereas \textit{S9} changes training data, model parameters, or trainable components.

\noindent \textbf{3) Taxonomy Characterization and Analysis.} We further characterize the assigned studies to compare LLM manipulation across the nine stages. \textbf{Manipulation Categories:} We group the nine stages into three broader categories based on the layer of attacker intervention.
a) Request \& Interaction Control covers \textit{S1--S5}, focusing on input- and interaction-level manipulation of LLM-related factors that affect task interpretation and generation.
b) Output Control covers \textit{S6--S7}, encompassing manipulation of output properties (\textit{S6}) and the victim-facing form or structure of phishing communication (\textit{S7}). 
c) Workflow \& Model Control covers \textit{S8--S9}, extending intervention to the system level across generation workflows, training data, and model components.

\textbf{Attack Methodology:} We identify Attack Tactics, Exploited AI Surface, Attack Goal, and Target LLMs in existing works to understand how attackers instantiate phishing generation methods, which vulnerabilities of LLMs are exploited (e.g., instruction-following tendencies~\cite{ouyang2022training}), what phishing goals are pursued, and to which LLMs the methods are applied. For instance, attackers set up an instruction \textit{``write a phishing email about...''} to frame the request as a writing task and steer the model toward drafting phishing content.

\textbf{Attack Properties:} We characterize attack properties through four aspects: a) Personalization Level (Per.) measures victim-specific tailoring in generated content ($\lowc$: low, $\medc$: medium, $\highc$: high). b) Automation Level (Aut.): captures the extent to which \textit{Attack Tactics} can be executed automatically after initial setup, where $\lowc$: largely manual prompting; $\medc$: semi-automated workflows with human guidance; $\highc$: highly automated generation or optimization pipelines. c) Attack Products (Prod.): denotes the generated phishing products: \faEnvelopeOpenText~= Phishing Email, \faPhone* = Vishing Script, \faLink~= URL Strings. d) Implementation Difficulty (Diff.): captures attacker effort and technical complexity, where \dA = direct prompt-based misuse with minimal setup; \dB = structured prompt engineering; \dC = tool-assisted or controlled generation; \dD = multimodal or agentic orchestration; and \dE = training- or optimization-intensive adaptation. Greater difficulty reflects more complex generation mechanisms, tools, or execution configurations.

\subsection{Existing Works Review}
This section reviews existing methods for LLM-enabled phishing generation across the nine stages, encompassing manipulation at the input, output, and system levels.

\textbf{1) Request \& Interaction Control --} \textit{\textbf{S1. Task Control}} Early attempts worked with direct, explicit commands (Per. = \lowc, Aut. = \lowc, Diff. = \dA). Attackers could simply prompt a model to \textit{``generate a bank transfer phishing email''}~\cite{eze2024analysis,ccetin2025exploring}, exploiting weak or immature safety guardrails. 
% Compared with ML and DL models, LLMs are more capable in both understanding and generation, leading to the emergence of LLM-based tools designed or adapted for phishing, such as WormGPT / FraudGPT~\cite{falade2023decoding}. 
Misuse of LLMs lowers the barrier to conducting phishing campaigns, making malicious content generation more accessible to low-skill attackers. These results highlight the vulnerability of under-aligned or self-hosted models to explicitly malicious instructions without requiring obfuscation or conversational setup.

\noindent \textit{\textbf{S2. Role \& Framing Control}} With platform-level refusals~\cite{open-policy}, attackers began masking intent through identity and role framing. Prompt templates asking the model to behave as a professional (for example, second person, \textit{``you are a cybersecurity expert''}~\cite{chen2024adapting}) or claiming legitimate research purposes (for example, first person, \textit{``I am a cybersecurity researcher''}~\cite{qi2025spearbot}) induce cooperation (Per. = \lowc, Aut. = \lowc, Diff. = \dA). These prompts exploit the models' tendency to help in seemingly authorized contexts rather than explicitly overriding safety rules. Although instruction-hierarchy schemes~\cite{wallace2024instruction} attempt to prioritize system-level constraints, role-framed prompts still remain effective. The harmful objective is not presented as an explicit policy violation, but as a task embedded in a plausible workflow. As a result, malicious tasks can be reframed as benign assistance.

\noindent \textit{\textbf{S3. Context Control}} Attackers can launch phishing campaigns by requesting LLMs to generate content adapted to a specific scenario. They can tailor prompts to particular target groups or organizations~\cite{weinz2025impact}, such as universities~\cite{karanjai2022targeted} and police departments~\cite{emanuela2024ai}. Attackers can also request phishing generation based on a given set of attributes, such as entities, URLs, and attachments~\cite{toth2026phishspamvalidgenerating}. These scenarios can be further refined with persuasion cues (e.g., \textit{Authority}) and generation rules (e.g., V-Triad) that increase perceived credibility (Per. = \medc, Aut. = \lowc, Diff. =\dB\cite{heiding2024devising,shim2024persuasion}). %One example combining GPT4 with V-Triad rules~\cite{heiding2024devising} to textit{``Create an email offering... for Harvard Students to Starbucks, with a link for them to access the QR code...''}, structures phishing content around \textit{credibility} (e.g., adding a logo of \textit{Starbucks}), \textit{compatibility} (e.g., aligning with the impersonated brand and target group), and \textit{customizability} (e.g., inserting scenario-specific QR code elements).
For example, one study combines GPT-4 with V-Triad rules in the prompt \textit{``Create an email offering... for Harvard Students to Starbucks, with a link for them to access the QR code...''}~\cite{heiding2024devising}. The prompt structures the phishing content around \textit{credibility}, \textit{compatibility}, and \textit{customizability}. 
Building on this, attackers can further move toward semi-automation by setting placeholders, \textit{``[principles here]''}, and asking LLMs to select or adapt suitable social engineering strategies for the given scenario (Per. = \medc, Aut. = \medc, Diff. = \dC\cite{bethany2025lateral,greco2024david}).

% \begin{tcolorbox}[
%   colback=gray!5, 
%   colframe=black!100, 
%   boxrule=0.3pt,
%   arc=0pt,
%   left=1mm,            
%   right=1mm,            
%   top=0.5mm,            
%   bottom=0.5mm          
% ]
% \textbf{Insight 3}: 
% Scenario-driven prompting turns simple text generation into modular attack templates. Effectiveness differs across LLM versions, and explicit labels such as ``phishing'' reduce transferability, prompting attackers to rely on implicit framing.
% %An (optional) mix of (phrases + attack scenarios + brand names + SE tactics, etc.) can expand LLM-phishing generation queries into attack templates. expands LLM-phishing generation queries into attack templates. It is varied in effectiveness across LLM versions due to differences in compliance mechanisms, and explicit labels in the template  (e.g., \textit{``phishing''}) tend to elicit refusals, undermining both cross-model transferability and long-term stealth~\cite{shyni2025generative}.
% \end{tcolorbox}

\noindent \textit{\textbf{S4. Target-Information Control}} Moving from scenarios to victim-level adaptation, attackers can retrieve victim-specific context from public social media and use it to guide phishing generation (Per. = \highc, Aut. = \medc, Diff. = \dB\cite{kim2025llms,czybik2026large,vafa2026context}). However, if collected profiles are outdated, such personalization may fail to capture the victim's current context. To further improve personalization, attackers can incorporate available signals from the physical environment (e.g., facial expressions and scene objects) and online activity (e.g., Instagram/LinkedIn posts) to adapt persuasive dialogues in live social engineering interactions~\cite{bi2025feasibility}. When victim-specific or real-time signals are unavailable, LLMs can instead synthesize plausible victim profiles to fill in missing details and maintain a coherent attack narrative, thereby reducing the need for manually collected victim information (Per. = \highc, Aut. = \highc, Diff. = \dC\cite{pajola2025phishgen,liu2025pimref}). Results~indicate that this synthetic personalization increases the deceptiveness of phishing content and can degrade existing ML/DL-based detectors~\cite{pajola2025phishgen}.

\begin{tcolorbox}[
  colback=gray!5, 
  colframe=black!100, 
  boxrule=1pt,
  arc=0pt,
  left=0mm,            
  right=0mm,            
  top=0.5mm,            
  bottom=0.5mm          
]
\textbf{Insight}: When real target information is incomplete or unavailable, LLMs can fabricate plausible victim-specific contexts and pretexts, enabling personalized-looking phishing content without corresponding real-world evidence. This shifts personalization from evidence-grounded adaptation toward synthetic context, but does not establish equivalent attack effectiveness. 
%\textbf{Insight}: LLMs reduce phishing personalization's dependence on collected victim information. Even when relevant target information is unavailable, LLMs can fabricate plausible victim-specific contexts, enabling personalized phishing without corresponding real-world evidence.

% LLM-assisted phishing personalization increasingly uses more than names, occupations, or organization membership. Recent methods retrieve social-media information, use current online or physical context, and generate plausible missing details when direct victim information is unavailable. The resulting phishing pretext can therefore reflect what a victim is currently doing or expecting, rather than only who the victim is.
\end{tcolorbox}

\begin{tcolorbox}[
  colback=gray!5, 
  colframe=black!100, 
  boxrule=1pt,
  arc=0pt,
  left=0.5mm,            
  right=0mm,            
  top=0mm,            
  bottom=0.5mm 
]
\textbf{Gap}: It remains unclear whether synthetic victim-specific context can match personalization grounded in real target information in inducing user action or evading detection. Controlled comparisons among no-target-information, synthetic-context, and real-target-information conditions are needed to determine whether LLMs actually reduce the information-gathering requirements of effective personalized phishing.
%\textbf{Gap}: It remains unclear whether fabricated victim-specific contexts can substitute for real target information in effective phishing personalization, and therefore whether LLMs meaningfully reduce the information-gathering requirements of personalized attacks.
\end{tcolorbox}

\noindent \textit{\textbf{S5. Interaction Control}} Beyond a single request, attackers may split a complete phishing request across multiple turns, each revealing part of the malicious intent (Per. = \lowc, Aut. = \lowc). The fragmented phishing intention usually starts with benign content (e.g., realistic email reply scenarios) or hypothetical statements, while subsequent prompts request sensitive information (e.g., payment details)~\cite{patel2023creatively,singh2023exploiting,shibli2024abusegpt}. This tactic sequentially builds phishing context across turns (Diff. = \dB) and makes each prompt conceal part of the malicious objective, reducing the effectiveness of single-round detection mechanisms and guiding the model toward final sensitive requests~\cite{yang2025chain,russinovich2025great}. 

% Singh et al.~\cite{singh2023exploiting} structure gradual dialogues that start with exploratory or hypothetical statements and only later request sensitive content. Patel et al.~\cite{patel2023creatively} and Shibli et al.~\cite{shibli2024abusegpt} leverage sequential prompt chaining within realistic email reply scenarios to invert the privacy-preserving behavior of LLMs. Early turns referenced benign content (flight confirmations, meeting summaries), while subsequent prompts requested sensitive information (payment details).

\noindent \textbf{2) Output Control --} \textit{\textbf{S6. Output-Property Control}} Beyond input-level manipulation, attackers can further manipulate properties of LLM-generated outputs. Existing methods rewrite phishing text (Aut. = \medc, Diff. = \dA\cite{afane2024next}), apply synonym substitution and sentence restructuring (Aut. = \medc, Diff. = \dA\cite{utaliyeva2023chatgpt}), or iteratively rephrase content with homoglyph substitution and polymorphic variations (Aut. = \medc, Diff. = \dB\cite{xue2025multiphishguard}). These transformations modify wording and surface features while preserving the underlying phishing intent, reducing explicit cues used by rule-based defenses.

\begin{tcolorbox}[
  colback=gray!5, 
  colframe=black!100, 
  boxrule=1pt,
  arc=0pt,
  left=0.5mm,            
  right=0.5mm,            
  top=0.5mm,            
  bottom=0.5mm          
]
\textbf{Insight}: LLM rewriting separates the phishing objective from the cues through which it is exposed, allowing attackers to preserve malicious intent while selectively controlling how explicit phishing cues are expressed and organized.
% Gap: Existing rewriting studies use different transformation constraints and source data, making it difficult to determine whether reported differences in phishing effectiveness or detector evasion arise from the rewriting mechanism itself or from the underlying content being rewritten.
\end{tcolorbox}

\noindent \textit{\textbf{S7. Communication Control}} Attackers can further manipulate the victim-facing communication structure of LLM-generated content. For instance, Vishing manipulates victim-facing conversational interactions by using LLMs to generate real-time responses that follow turn-taking, handle victim reactions, and maintain a chosen persona, tone, and conversational style (Per. = \highc, Aut. = \highc, Diff. = \dD~\cite{figueiredo2025sounds,toapanta2024ai}). Generated dialogues may also incorporate benign conversational noise (e.g., small talk) to reduce the salience of malicious intent (Per. = \lowc~\cite{li2025talking}). In experiments with LLM-enhanced personalized scripts and real-time dialogue, about half of victims disclosed sensitive information, including one-third despite a clear warning~\cite{figueiredo2025sounds}. These interactions challenge timely risk recognition and call for defenses against adaptive conversational phishing.
% Combined with LLM-enhanced personalized scripts and real-time dialogue, these attacks make malicious intent harder to recognize and increase disclosure success~\cite{figueiredo2025sounds}.

Communication control can also target victim-facing actionable artifacts (e.g., URLs). In Quishing, phishing URL strings are concealed behind QR codes rather than directly exposed in phishing messages~\cite{roy2024chatbots}. Weinz et al.~\cite{weinz2025impact} show that QR-based messages achieve user engagement comparable to directly displayed links despite requiring an additional scanning step. The change conceals malicious URL strings from direct inspection and introduces visual cues that make phishing risks harder for users to assess before interaction~\cite{kowalewski2025scanned}.

\noindent \textbf{3) Workflow \& Model Control --} \textit{\textbf{S8. Workflow Control}} Extending from manual prompting, attackers can organize LLM generation into multi-step workflows with limited human intervention. Roy et al.~\cite{roy2024chatbots} show that LLMs can generate malicious prompt variants that are subsequently fed back into the model to produce additional phishing attacks, substantially reducing manual prompt-engineering effort. Even simple instructions, such as \textit{``summarize the given content,''} can initiate this process and produce multiple safety-bypassing variants (Aut. = \highc, Diff. = \dB~\cite{Codebook}). The workflow can further incorporate strategy selection before generation (Diff. = \dE~\cite{khan2021offensive}), followed by iterative generation, evaluation, and selection of stronger variants (Diff. = \dE~\cite{fairbanks2024generating}). Tewari et al.~\cite{tewari2026trust} further extend such workflows into an agentic, largely automated phishing pipeline that coordinates multi-turn attacker--victim interactions with downstream action verification. These manipulation mechanisms extend attacker intervention from individual prompts to the generation workflow, enabling repeated prompt generation and phishing-content production at scale.

\begin{tcolorbox}[
  colback=gray!5, 
  colframe=black!100, 
  boxrule=1pt,
  arc=0pt,
  left=0.5mm,            
  right=0mm,            
  top=0mm,            
  bottom=0.5mm          
]
\textbf{Gap}: Existing studies establish the feasibility and throughput of automated phishing generation, but rarely compare automated workflows with matched single-pass or manually guided baselines. It therefore remains unclear whether workflow automation, beyond producing more variants, improves phishing quality, user compliance, or detector evasion.
%\textbf{Gap}: Existing studies establish that phishing generation can be automated, but provide limited evidence on whether automation improves attack quality, user compliance, or detector evasion beyond increasing generation volume.
% Automation changes how phishing content is produced, but automation alone does not establish a stronger phishing attack. Its security value depends on whether iterative generation, evaluation, and selection improve attack outcomes beyond simply increasing the number of generated variants. Existing workflow studies demonstrate that automated phishing generation is feasible, but provide limited evidence on whether automation improves phishing quality, user compliance, or detector evasion beyond increasing generation throughput.
\end{tcolorbox}

\noindent \textit{\textbf{S9. Model Control}} At the model level, attackers can directly modify the generation model through task-specific training or adaptation. Existing approaches train phishing-oriented neural language models~\cite{guo2022generating}, incorporate trainable adapters for cross-lingual phishing generation~\cite{guo2024x}, or fine-tune LLMs on phishing data~\cite{mehdi2023adversarial,panda2024teach}. Model adaptation can further involve adversarial training, where the generator is iteratively updated against a discriminator to improve phishing realism~\cite{chen2024adapting}. These methods modify learned model parameters or components, but require dedicated training data and greater implementation effort (Diff. = \dE).

\begin{tcolorbox}[
  colback=gray!5, 
  colframe=black!100, 
  boxrule=1pt,
  arc=0pt,
  left=0mm,            
  right=0mm,            
  top=0.5mm,            
  bottom=0.5mm          
]
\textbf{Gap}: Across the nine stages, phishing quality and attack effectiveness remain insufficiently distinguished.Existing studies mainly assess whether LLM outputs resemble plausible phishing, while fewer measure attacker objectives such as clicks, disclosure, or payment. Consequently, evidence of fluent or convincing generation should not be treated as evidence of effective phishing.
%\textbf{Gap} Phishing quality and phishing effectiveness are distinct: the former characterizes the generated content, whereas the latter reflects whether it achieves the intended phishing objective. Existing studies primarily examine how LLMs can generate phishing content, with some further evaluating its phishing quality, while providing limited evidence on whether the generated outputs actually constitute effective phishing rather than merely phishing-like content.
\end{tcolorbox}
% \subsection{Threats In-the-Wild}
% LLM capabilities have been repackaged into commercialized phishing tools and services~\cite{Trend_2023}. One representative example is WormGPT~\cite{falade2023decoding}, promoted in 2023 as a GPT-J-based phishing tool and estimated to have generated over \$28,000 in revenue within roughly two months~\cite{lin2024malla}. These services rarely depend on fundamentally new model development; instead, they often use jailbreak-as-a-service to bypass safeguards at the prompt level~\cite{Trend_2026,trend_surging_hype}, lowering the barrier to malicious adoption (e.g., KawaiiGPT is configurable in under five minutes~\cite{unit42,kawaii_gpt}). Further, LLM-misuse expands from phishing email generation to advertise code obfuscation, cookie/log replication~\cite{csico_talos}, integrating LLM APIs into malware and broader attack chains, enabling more automated and adaptive attacks~\cite{mandiant_ai_risk,Trend_2026}. 

\begin{table*}[!ht]
  \caption{Systematization of LLM-generated phishing attributes and their user-side effects.}
  \label{tab:llm_phish_attr_overview}
  \centering
  \scriptsize
  \setlength{\tabcolsep}{1pt}
  \renewcommand{\arraystretch}{0.8}

  \resizebox{\linewidth}{!}{
  \begin{tabular}{
    L{1.2cm}
    L{1.9cm}
    L{1.3cm}
    L{0.7cm}
    L{1.7cm}
    C{1.2cm}
    L{1cm}
    L{5cm}
    C{1.1cm}
    L{6.5cm}
  }
    \toprule
    \multicolumn{1}{c}{\multirow[c]{3}{*}{\makecell[c]{\textbf{Paradigms}}}}&
    \multicolumn{1}{c}{\multirow[c]{3}{*}{\makecell[c]{\textbf{Objects}}}} &
    \multicolumn{6}{c}{\textbf{Analytical Perspective}} &
    \multicolumn{2}{c}{\textbf{User-Side Effects}} \\

    \cmidrule(r){3-8} \cmidrule(r){9-10}
    & &
    \multirow[c]{1}{*}{\makecell[c]{\textbf{Example}}} &
    \multirow[c]{1}{*}{\makecell[c]{\textbf{Works}}} &
    \shortstack[c]{\textbf{Data Source}} &
    \shortstack[c]{\textbf{Generation}\\\textbf{Basis}} &
    \shortstack[c]{\textbf{Phishing}\\\textbf{Eval.}} &
    \textbf{Research Focus} &
    \shortstack[c]{\textbf{Impact}\\\textbf{Type}} &
    \textbf{Reported Effect} \\
    \midrule

% ===================== Text Traits =====================
\multicolumn{1}{c}{\multirow[c]{10}{*}{\makecell[c]{Text\\Traits} }}
& \multicolumn{1}{c}{\multirow[c]{6}{*}{\makecell[c]{Textual\\Characteristics}}}
& \multicolumn{1}{c}{\multirow[c]{6}{*}{\makecell[c]{lexical,\\syntactic,\\fluency}}}
& \cite{afane2024next}
& HW-E,LLM-A
& S6
& \lowc
& phishing-variant generation for detector evaluation
& Exp.
& -- \\

& 
& 
& \cite{heiding2024devising}
& \shortstack[l]{HW-E/A, LLM-A}
& S3
& \lowc
& phishing effectiveness and persuasion
& Beh.
& Higher click tendency \\

& 
& 
& \cite{eze2024analysis}
& HW-E,LLM-A
& S1
& \highc
& content realism and linguistic differences
& Exp.
& -- \\

& 
& 
& \cite{hao2025spammers}
& \shortstack[l]{HW-E,LLM-A}
& S3
& \lowc
& language sophistication evolution
& Exp.
& -- \\

& 
& 
& \cite{veisi2025user}
& N/A
& N/A 
& \lowc
& user distinguishability
& Attr.
& Superficial cues support bot detection \\

& 
& 
& \cite{malloy2025improving}
& HW-E,LLM-A
& S6
& \lowc
& user distinguishability
& Attr.
& Human-written, LLM-stylized emails were hardest to identify \\

\cmidrule(r){2-10}

& \multicolumn{1}{c}{\multirow[c]{3}{*}{\makecell[c]{Social\\Engineering\\Tactics}}}
& \multicolumn{1}{c}{\multirow[c]{3}{*}{\makecell[c]{authority,\\scarcity}}}
& \cite{emanuela2024ai}
& LLM-A
& S3
& \lowc
& AI-assisted social engineering
& Mod.
& -- \\

&
&
& \cite{chen2024adapting}
& HW-E,LLM-A
& S9
& \highc
& evolution of phishing strategies
& Mod.
& -- \\

&
&
& \cite{popovic2026dideco}
& LLM-A
& N/A
& \lowc
& phishing-intent structure
& --
& Concurrent explicit goals and implicit persuasion mechanisms \\

\midrule

% ===================== Human Factors =====================
\multicolumn{1}{c}{\multirow[c]{5}{*}{\makecell[c]{Human\\Factors} }}
% \multicolumn{1}{c}{\multirow[c]{4}{*}{\makecell[c]{Model\\Adapted}}}
& \multicolumn{1}{c}{\multirow[c]{3}{*}{\makecell[c]{Individual\\Characteristics}}}
& \multicolumn{1}{c}{\multirow[c]{3}{*}{\makecell[c]{demographics,\\education}}}
& \cite{alahmed2024exploring}
& N/A
& N/A
& \lowc
& targeting realism and personalization
& Mod.
& Broad acceptance across occupations \\

&
&
& \cite{francia2024assessing}
& HW-A,LLM-A
& S4
&\lowc
& user perception and response
& Leg.
& Higher persuasiveness and poor human-vs-AI source attribution \\

&
&
& \cite{olea2025evaluating}
& \shortstack[l]{HW-E,LLM-A}
&  N/A
& \lowc
& detection accuracy experiment
& Attr.
& Lower human-vs-AI source attribution accuracy \\

\cmidrule(r){2-10}

& \multicolumn{1}{c}{\multirow[c]{2}{*}{\makecell[c]{Psychological\\Characteristics}}}
& \multicolumn{1}{c}{\multirow[c]{2}{*}{\makecell[c]{personality,\\cognitive bias}}}
& \cite{asfour2023harnessing}
& HW-E,LLM-A
& N/A
& \lowc
& simulated victim response
& Mod.
& Agreeable and less conscientious traits increase susceptibility \\

&
&
& \cite{sharma2023well}
& HW-A,LLM-A
& S3
& \highc
& bias-aware phishing effectiveness
& Mod.
& Human-crafted emails showed higher phishing effectiveness \\

\midrule

% ===================== Model Traits =====================
\multicolumn{1}{c}{\multirow[c]{1}{*}{\makecell[c]{Model\\Traits}}}
& \multicolumn{1}{c}{\multirow[c]{1}{*}{\makecell[c]{Scalability}}}
& \multicolumn{1}{c}{\multirow[c]{1}{*}{\makecell[c]{speed, cost}}}
& \cite{schmitt2024digital}
& N/A
& N/A
& \lowc
& GenAI-enabled phishing shift
& Exp.
& Increased attack volume and realism \\

& \multicolumn{1}{c}{\multirow[c]{1}{*}{\makecell[c]{Cross-Model Variation}}}
& \multicolumn{1}{c}{\multirow[c]{1}{*}{\makecell[c]{output diff.}}}
& \cite{vafa2026context}
& HW-E,LLM-A
& S4
& \highc
& cross-model comparison with real-world phishing
& Leg.
& Lower human suspicion \\

\bottomrule
  \end{tabular}
  }
\end{table*}

\section{Characterizing LLM-Generated Phishing}\label{RQ2}
% \noindent To address RQ2, we examine 14 studies that describe how LLM-generated phishing differs from human-written attacks across three perspectives: textual traits, human factors manipulation, and model-level properties. Collectively, these works suggest that LLM-generated phishing does not introduce entirely new attack forms. Instead, it modifies the quality and structure of existing techniques by smoothing recognizable linguistic patterns, amplifying social engineering strategies, and aligning narratives more closely with contextual cues such as user profiles. Table~\ref{tab:llm_phish_attr_overview} summarizes representative studies mapped to each perspective.

\subsection{Systematization Methodology}
To understand how LLM-generated phishing differs from human-written phishing, we organize existing studies along three analytical paradigms (Table~\ref{tab:llm_phish_attr_overview}):
\textbf{Analysis Paradigms} refer to the level at which phishing content and its effects are examined. a) Text Traits focus on textual properties and social engineering patterns; b) Human Factors examine user-related attributes such as demographics and psychological traits to understand how different users perceive and respond to synthesized phishing; and c) Model Traits consider the capability and efficiency, highlighting how scaling and automation affect the phishing threat landscape.

\textbf{Analytical Perspective} captures how prior work designs and supports the analysis. a) Comparison Dimension specifies what is being compared; b) Data Source indicates whether phishing samples are human-written (HW) or LLM-generated (LLM), and whether they come from existing datasets (E) or are created by the authors (A), yielding HW-E, HW-A, LLM-E, and LLM-A; c) Generation Basis denotes the (\S\ref{RQ1}) stage used to generate the analyzed samples; d) Phishing Eval. indicates whether LLM-generated phishing samples are evaluated; \textit{N/A} denotes the evaluation is not applicable; and e) Research Focus reflects the goal of the study, such as measuring persuasion effectiveness;
% \textbf{c) Evidence Basis} indicates that the conclusions are supported via behavioral experiments (Beh.), perception studies (Perc.), simulations (Sim.), qualitative analysis (Qual.), and system-level observations (Sys.).

\textbf{User-Side Effects:} a) Impact Type captures the forms through which LLM-generated phishing influences users, including increased exposure to phishing content (Exp.), heightened perceived legitimacy (Leg.), strengthened behavioral susceptibility (Beh.), impaired source attribution (Attr.), and user-dependent differences in susceptibility (Mod.). b) Reported Effects demonstrate the representative impact of LLM-generated phishing attacks on users. 
% Overall, Table~\ref{tab:llm_phish_attr_overview} indicates that LLM-generated phishing affects users across exposure, conception, action, and attribution, making anti-phishing defense more challenging.

\subsection{Existing Works Review}
\textbf{Textual Traits:} 
% Across empirical studies, LLM-generated phishing is generally more fluent, coherent, and conversational than human-written phishing, which more often contains rigid wording and linguistic errors.
Prior analyses report that LLM-generated phishing has more verbs and pronouns but fewer numeric tokens, suggesting greater emphasis on actions and interpersonal engagement~\cite{eze2024analysis}. LLM-generated phishing can also evade detection by reducing overt grammatical and spelling errors while preserving subtle imperfections that maintain authenticity~\cite{malloy2025improving,hao2025spammers,veisi2025user}. It can further exhibit benign-like characteristics, including part-of-speech usage, lexical diversity, and sentence-length patterns closer to legitimate business writing~\cite{afane2024next}. These properties make LLM-generated phishing less distinguishable from routine content using conventional phishing features~\cite{hao2025spammers,eze2024analysis}.

Existing works show that LLM-generated phishing often combines a broader set of persuasion principles (e.g., \textit{Scarcity}, \textit{Authority}, and \textit{Consistency})~\cite{chen2024adapting,popovic2026dideco}, whereas human-written phishing relies more heavily on a smaller set (e.g., \textit{Authority} and \textit{Scarcity})~\cite{chen2024adapting}. User detection studies further show that scarcity-based messages are recognized more easily, while adding authority significantly reduces detection accuracy~\cite{emanuela2024ai}. The combined use of multiple persuasion principles increases the complexity of phishing strategies, making malicious intent harder for users to identify.

\begin{tcolorbox}[
  colback=gray!5, 
  colframe=black!100, 
  boxrule=1pt,
  arc=0pt,
  left=0.5mm,            
  right=0mm,            
  top=0mm,            
  bottom=0.5mm          
]
\textbf{Gap}: Existing characterization emphasizes lexical, stylistic, and persuasion features, although LLMs can alter several simultaneously. Similar features also appear in benign content and do not directly indicate phishing intent or malicious action. Evidence remains limited on how these LLM-induced changes relate to phishing intent, requested actions, and downstream detection.
% Similar properties also appear in benign content and do not directly indicate phishing intent or malicious action. However, there is limited evidence on how LLM-induced changes in these properties relate to the expression of phishing intent and requested actions, or how these relationships affect downstream detection.
\end{tcolorbox}

\textbf{Human Factors:} 
Studies on human factors show that susceptibility to LLM-generated phishing varies across users. Lower technical literacy and limited security training increase vulnerability, yet convincing scenarios that impersonate authority figures can still deceive experienced users~\cite{alahmed2024exploring}. 
Users across age groups may perceive more credible LLM-generated persona contexts as more trustworthy and show greater willingness to comply with malicious requests~\cite{francia2024assessing}.
Psychological factors, such as overconfidence, susceptibility to persuasion, and curiosity, may reduce users' attention to phishing cues while increasing the perceived legitimacy of deceptive content~\cite{sharma2023well,asfour2023harnessing,olea2025evaluating}. These findings show that phishing outcomes are influenced by multiple factors related to both users and generated phishing content.
\begin{tcolorbox}[
  colback=gray!5, 
  colframe=black!100, 
  boxrule=1pt,
  arc=0pt,
  left=0.5mm,            
  right=0mm,            
  top=0mm,            
  bottom=0.5mm          
]
\textbf{Gap}: Existing user studies often stop at AI-authorship, credibility, or phishing-recognition judgments, which do not establish whether users avoid the attacker’s intended action. Identifying AI authorship does not imply recognizing deceptive intent, nor does recognizing phishing necessarily prevent clicking, disclosure, or payment. Evidence linking these judgments to actual behavior remains limited.
%\textbf{Gap}: AI-authorship attribution is distinct from phishing recognition and user compliance. Recognizing that a message is AI-generated does not reveal whether users perceive its deceptive intent, while recognizing phishing does not necessarily prevent the intended action.
% AI-authorship attribution, phishing recognition, perceived legitimacy, and harmful user action are distinct user outcomes. Correctly identifying a message as AI-generated does not establish that a user recognizes phishing, and recognizing phishing does not necessarily prevent compliance. Many existing user studies stop at authorship judgments, perceived credibility, or phishing-recognition decisions. Evidence about whether users can distinguish AI-written messages therefore provides limited evidence about actual security outcomes such as clicking, disclosing information, authenticating, or making payments. 
\end{tcolorbox}

\textbf{Model Traits:} LLMs affect phishing not only through message content, but also through how attacks can be produced and scaled. Personalized phishing can be generated 96\% faster than human-written attempts while more than doubling user engagement~\cite{schmitt2024digital}. Moreover, both generation efficiency and resulting phishing characteristics can vary substantially across model types. Vafa et al.~\cite{vafa2026context} find that commercial LLMs generally produce more persuasive and credible phishing, while open-source LLMs show greater variation in persuasion, sender credibility, and linguistic naturalness. These findings highlight the need to consider model differences when analyzing LLM-generated phishing.

\subsection{Persuasion Patterns across Communication Settings}\label{rq2_industry_discussion}
In this section, we conduct an exploratory comparison of persuasion patterns across LLM-generated phishing communication settings. Because these settings are represented by different public datasets and generation procedures, the observed differences should not be attributed to communication setting alone. 
%In this section, we discuss differences in persuasion patterns across LLM-generated phishing modalities. 
% Because the available datasets and extracted features are text-based, we compare email and Vishing payloads here and discuss Quishing detector performance separately in Appendix~\ref{appendix:quishing_bench}. 
We use persuasion principles~\cite{cialdini2001science}, which appear in both academic and industry guidance~\cite{microsoft_anti_phishing,google_anti_phishing,proofprint_anti_phishing,mimecast_anti_phishing}, as our analytical framework. %We adopt a general persuasion principle annotation method~\cite{chen2021weakly} to avoid biased annotation on specific datasets or attack vectors.
We adopt a general persuasion-principle annotation method~\cite{chen2021weakly} to apply a consistent annotation procedure across datasets and attack vectors. Our datasets consist of recent publicly available resources; details are available in Appendix~\ref{appendix:benchmarking_datasets}. 
% \tina{Fengchao: Please report the sample count for each communication setting, the unit of analysis, the overall and pairwise statistical tests used, and the FDR correction family. Because the three settings come from different datasets and generation procedures, please also confirm that the results are presented as an exploratory cross-dataset comparison rather than as a causal effect of communication setting.}
% \begin{tcolorbox}[
%   colback=gray!5, 
%   colframe=black!100, 
%   boxrule=1pt,
%   arc=0pt,
%   left=0.5mm,            
%   right=0mm,            
%   top=0mm,            
%   bottom=0.5mm          
% ]
% \textbf{Gap}: \fc{The compared communication settings originate from different datasets and generation procedures, so the observed differences cannot be attributed to communication setting alone.} It also remains unclear how persuasion progression changes in response to victim behavior. %我们的正文阐述要注意阐明这一点。只是分析特征变化，communication的不同，特征也确实不同。 要说清楚最开始比的是什么，有没有考虑datasets的问题。
% \end{tcolorbox}

\begin{figure}[!ht]
    \centering
    \includegraphics[width=\linewidth]{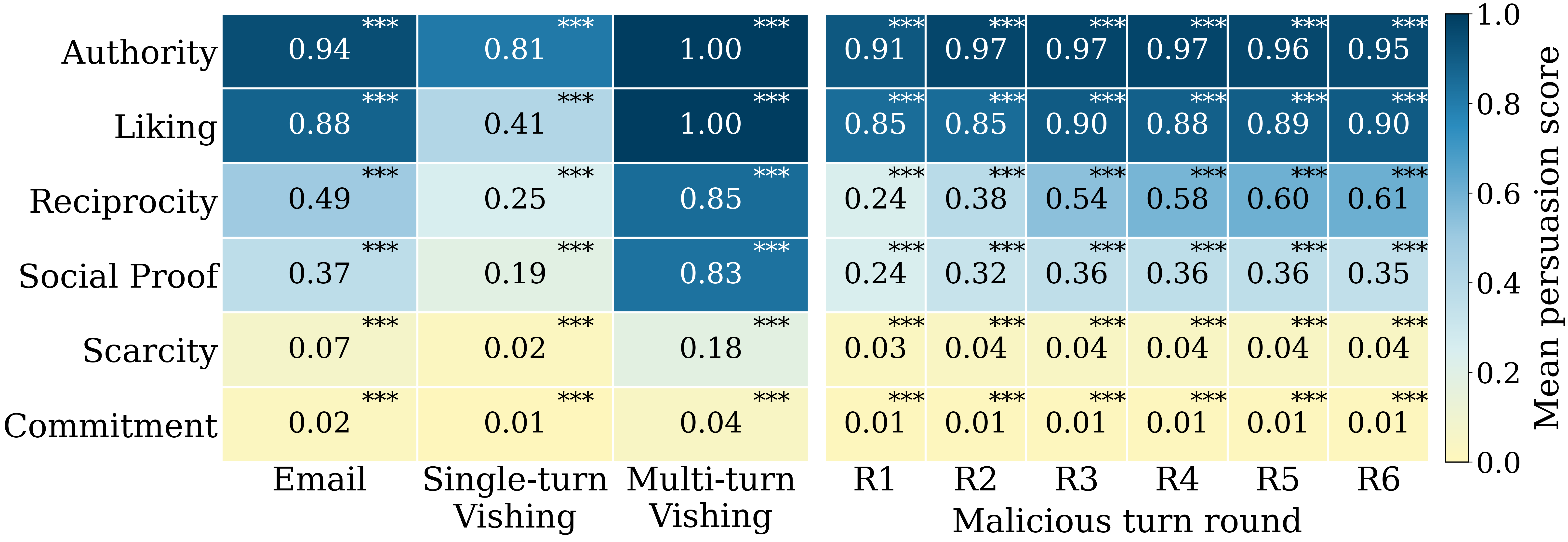}%,trim=0.7cm 0.3cm 0.7cm 0.3cm, clip
    \caption{Persuasion patterns across communication settings in LLM-generated phishing. The left panel compares mean persuasion scores across email, single-turn Vishing, and multi-turn Vishing, while the right panel shows how these principles evolve over the first six malicious turns in multi-turn Vishing.}
    \label{fig:vishing_multi_patterns}
\end{figure}
We compare persuasion patterns across LLM-generated phishing datasets covering email, single-turn Vishing, and multi-turn Vishing using the same six persuasion principles (Fig.~\ref{fig:vishing_multi_patterns}). All six principles differ significantly across the three communication settings after FDR correction (all $q<0.001$). Multi-turn Vishing exhibits the highest mean scores across all six principles, whereas single-turn Vishing consistently exhibits the lowest. The largest differences occur in \textit{Social Proof}, \textit{Reciprocity}, and \textit{Liking}: compared with single-turn Vishing, their scores increase from 0.19 to 0.83, 0.25 to 0.85, and 0.41 to 1.00 in multi-turn Vishing, respectively, with all pairwise differences significant ($q<0.001$). These relational cues are also among the persuasion strategies that prior comparisons find less emphasized in human-written phishing, which relies more heavily on \textit{Authority} and \textit{Scarcity}; \textit{Scarcity} is still present in our LLM-generated datasets, but at relatively low scores.

These differences are also reflected in how persuasion unfolds across communication settings. Emails can combine multiple persuasion strategies within a single static message, whereas multi-turn Vishing can distribute them across successive dialogue rounds. \textit{Authority} and \textit{Liking} remain consistently high across the first six malicious turns (0.91--0.97 and 0.85--0.90), supporting early trust establishment and attention maintenance. In contrast, \textit{Reciprocity} increases from 0.24 in the first turn to 0.61 in the sixth, becoming particularly visible from the third round onward through offers of assistance or solutions (e.g., ``transfer to supervisor'') that encourage continued engagement. \textit{Social Proof} similarly rises early and then stabilizes. %These patterns suggest that Vishing can progressively introduce persuasion as interaction develops, whereas email must concentrate persuasive strategies within a single message, creating communication-setting-specific patterns that defenses need to account for.
Within the analyzed datasets, these patterns are consistent with multi-turn Vishing distributing persuasion across successive turns, whereas email presents persuasive strategies within a single message. Because the datasets and generation procedures differ, the analysis identifies cross-setting patterns but does not isolate a causal effect of communication setting.

\begin{tcolorbox}[
  colback=gray!5, 
  colframe=black!100, 
  boxrule=1pt,
  arc=0pt,
  left=0.5mm,            
  right=0mm,            
  top=0mm,            
  bottom=0.5mm          
]
\textbf{Gap}: Existing evidence does not isolate communication setting from dataset, generator, and generation-procedure differences. It also remains unclear how LLMs adapt manipulation strategies to actual victim responses during multi-turn phishing and whether such response-conditioned adaptation changes user behavior. Controlled studies should hold the attack goal, target information, generator, and generation instructions constant while varying communication settings and victim responses.
%\textbf{Gap}: Existing evidence does not isolate communication setting from dataset, generator, and generation-procedure differences. It also remains unclear whether LLMs adapt manipulation to victim responses and how adaptation affects behavior. Controlled studies should hold attack goal, target information, generator, and instructions constant while varying communication settings and victim responses.
%\textbf{Gap}: Persuasion differs across communication settings not only in prevalence but also in how it unfolds. However, it remains unclear how LLMs adapt manipulation strategies to victim responses during multi-turn phishing, and whether such response-conditioned adaptation affects user behavior or the likelihood of inducing the intended action.
% Existing evidence across communication settings comes from different datasets and generation procedures, which limits causal interpretation. Future evaluations should hold the attack goal, target information, generator, and generation instructions constant while changing the communication setting. Controlled comparisons would show which phishing characteristics arise from the communication setting itself rather than from dataset or generator differences.
\end{tcolorbox}

\begin{table*}[!t]
  \caption{Systematization of defenses against LLM-generated phishing content.}
  \label{tab:detection_methods_differ_rq3}
  \centering
  \scriptsize
  \setlength{\tabcolsep}{0.8pt}
  \renewcommand{\arraystretch}{0.5}
  \begin{threeparttable}

  \resizebox{\linewidth}{!}{
  \begin{tabular}{
    p{0.6cm}
    p{1.5cm}
    p{0.5cm}
    p{0.5cm}
    p{0.8cm}
    c
    c
    p{3cm}
    p{1cm}
    C{0.5cm}
    C{0.5cm}
    C{0.5cm}
    c
    c
    c
    c
    c
    c
  }
    \toprule
    \multicolumn{1}{c}{\multirow{2}{*}{\makecell[c]{\textbf{Paradigms}}}}
      & \multicolumn{1}{c}{\multirow{2}{*}{\makecell[c]{\textbf{Objects}}}}
      & \multicolumn{3}{c}{\textbf{Defense Scope}}
      & \multirow{2}{*}{\textbf{Work}}
      & \multicolumn{3}{c}{\textbf{Defense Methodology}}
      & \multicolumn{3}{c}{\makebox[0pt][c]{\textbf{Input Requirements}}}
      & \multicolumn{6}{c}{\textbf{Operational Properties}} \\

    \cmidrule(l){3-5}
    \cmidrule(lr){7-9}
    \cmidrule(lr){10-12}
    \cmidrule{13-18}

      &
      &
      \textbf{Goal}
      & \textbf{Plc.}
      & \textbf{Stg.}
      &
      & \textbf{App.}
      & \textbf{Defense Model}
      & \textbf{Evd.}
      & \textbf{H}
      & \textbf{C}
      & \textbf{U}
      & \textbf{Data}
      & \textbf{Prompt}
      & \textbf{XGen}
      & \textbf{Art.}
      & \textbf{Ext.}
      & \textbf{Repro.} \\

    \midrule

    \multicolumn{1}{c}{\multirow[c]{25}{*}{\makecell[c]{Content\\Tailored\\Defense}}}
      & \multicolumn{1}{c}{\multirow[c]{7}{*}{\makecell[c]{Textual\\Characteristics\\Screening}}}
      & \faEnvelope & O & \textit{S1,S3}
      & \cite{elongha2024detecting}
      & ML/DL
      & RF/CNN + NN, etc.
      & ST
      & -- & \checkmark & --
      & \highc & \lowc & \lowc & \lowc & \lowc
      & $\blacksquare\blacksquare\square\square\square$ \\
      
      & & \faPenNib & O & \textit{N/A}
      & \cite{gryka2024detection}
      & ML
      & RF, SVM, MLP, kNN, etc.
      & ST
      & -- & \checkmark & --
      & \highc & \lowc & \lowc & \lowc & \lowc
      & $\blacksquare\blacksquare\square\square\square$ \\
      
      & & \faPenNib & O & \textit{N/A}
      & \cite{wani2024ai}
      & DL
      & CNN, GRU, BiLSTM, etc.
      & ST
      & -- & \checkmark & --
      & \highc & \lowc & \lowc & \lowc & \lowc
      & $\blacksquare\blacksquare\blacksquare\square\square$ \\ 
     
      & & \faPenNib & O & \textit{S1}
      & \cite{eze2024analysis}
      & ML/DL
      & MALLET, WNN, LSTM
      & ST
      & -- & \checkmark & --
      & \highc & \lowc & \lowc & \highc & \lowc
      & $\blacksquare\blacksquare\blacksquare\square\square$ \\
     
      & & \faPenNib & O & \textit{S1,S3}
      & \cite{brissett2025machine}
      & ML
      & LR
      & ST+TAG
      & -- & \checkmark & --
      & \highc & \lowc & \lowc & \medc & \lowc
      & $\blacksquare\blacksquare\blacksquare\square\square$ \\

      & & \faPenNib & O & \textit{S1,S3}
      & \cite{opara2025evaluating}
      & ML
      & XGBoost, LR, RF, etc.
      & ST
      & -- & \checkmark & --
      & \highc & \lowc & \lowc & \medc & \lowc
      & $\blacksquare\blacksquare\blacksquare\square\square$ \\
      
      & & \faEnvelope & O & \textit{N/A}
      & \cite{malik2026qfedphish}
      & FL
      & Quantum-inspired classifier
      & ST
      & -- & \checkmark & --
      & \highc & \lowc & \lowc & \lowc & \highc
      & $\blacksquare\blacksquare\square\square\square$ \\

    \cline{2-18}
    
      & \multicolumn{1}{c}{\multirow[c]{4}{*}{\makecell[c]{Social\\Engineering\\Modeling}}}
      & \faEnvelope & O & \textit{S1,S3}
      & \cite{heiding2024devising}
      & LLMs
      & Claude, ChatGPT, Bard, etc.
      & SE + ST
      & -- & \checkmark & --
      & \lowc & \highc & \highc & \lowc & \highc
      & $\blacksquare\blacksquare\square\square\square$ \\
     
      & & \faEnvelope & O & \textit{S1,S4}
      & \cite{nahmias2024prompted}
      & LLMs/ML
      & GPT-3.5, Gemini + kNN
      & SE
      & -- & \checkmark & --
      & \medc & \medc & \lowc & \highc & \highc
      & $\blacksquare\blacksquare\blacksquare\blacksquare\square$ \\
            
      & & \faEnvelope & O & \textit{S1,S6}
      & \cite{xue2025multiphishguard}
      & LLMs/RL
      & GPT-4o + PPO
      & SE + ST
      & \checkmark & \checkmark & \checkmark
      & \medc & \medc & \lowc & \medc & \highc
      & $\blacksquare\blacksquare\blacksquare\square\square$ \\
      
      & & \faPenNib & O & \textit{N/A}
      & \cite{bethany2025lateral}
      & DL
      & T5 encoder + MLP
      & SE + ST
      & -- & \checkmark & --
      & \medc & \lowc & \highc & \lowc & \highc
      & $\blacksquare\blacksquare\blacksquare\blacksquare\square$ \\

    \cline{2-18}

      & \multicolumn{1}{c}{\multirow[c]{3}{*}{\makecell[c]{Intention\\Screening}}}
      & \faBullseye & P & \textit{S1,S8}
      & \cite{roy2024chatbots}
      & DL
      & RoBERTa, DeBERTa, etc.
      & PR
      & -- & \checkmark & --
      & \highc & \lowc & \highc & \medc & \lowc
      & $\blacksquare\blacksquare\blacksquare\blacksquare\square$ \\

      & & \faBullseye & P/O & \textit{S1,S9}
      & \cite{pang2025paladin}
      & LLMs
      & Instrumented LLaMA/Qwen
      & TAG
      & -- & \checkmark & --
      & \highc & \lowc & \lowc & \highc & \highc
      & $\blacksquare\blacksquare\blacksquare\blacksquare\square$ \\

      & & \faBullseye & P & \textit{S1,S4,S5}
      & \cite{vafa2026context}
      & DL/LLMs
      & RoBERTa, LlamaGuard-3, etc.
      & PR + POL
      & -- & \checkmark & --
      & \medc & \lowc & \lowc & \medc & \lowc
      & $\blacksquare\blacksquare\blacksquare\square\square$ \\

    \cline{2-18}

      & \multicolumn{1}{c}{
          \multirow[c]{3}{*}{
            {\makecell[c]{Rule\\Compliance\\Screening}}
          }
        }
      & \faEnvelope & O & \textit{S1,S3}
      & \cite{mahendru2024securenet}
      & DL/LLMs
      & DeBERTa-v3, Gemini, etc.
      & ST + SE
      & -- & \checkmark & \checkmark
      & \medc & \medc & \highc & \medc & \highc
      & $\blacksquare\blacksquare\blacksquare\square\square$ \\

      & & \faEnvelope/\faUser & O/U & N/A
      & \cite{quinn2024applying}
      & LLMs
      & Gemini
      & POL
      & -- & \checkmark & --
      & \medc & \highc & \lowc & \medc & \highc
      & $\blacksquare\square\square\square\square$ \\

      & & \faEnvelope & O & \textit{S1,S4}
      & \cite{liu2025pimref}
      & DL
      & NER + CharacterBERT
      & KB 
      & \checkmark & \checkmark & --
      & \lowc & \highc & \lowc & \medc & \medc
      & $\blacksquare\blacksquare\blacksquare\blacksquare\blacksquare$ \\

    \midrule

    \multicolumn{1}{c}{\makecell[c]{Human-Centric\\Defense}}
      & \multicolumn{1}{c}{\makecell[c]{Behavioral\\Analysis}}
      & \faUser & U & \textit{S1}
      & \cite{malloy2025training}
      & Cog./LLMs
      & IBL + GPT-4
      & BH
      & -- & \checkmark & --
      & \highc & \lowc & \lowc & \lowc & \medc
      & $\blacksquare\square\square\square\square$ \\

    \bottomrule
  \end{tabular}
  }

  \begin{tablenotes}[flushleft]
  \footnotesize
  \item \textbf{Defense Evidence (Evd.)} records the information used by a defense for detection or intervention, including textual/stylometric evidence (ST), social-engineering cues (SE), prompt/request evidence (PR), external knowledge (KB), trigger/tag evidence (TAG), security policy (POL), and user behavioral evidence (BH). \textbf{Input Requirements:} email headers (H), phishing communication text (C), and URL strings (U). Communication text includes email, SMS bodies, and Vishing scripts or messages. $\lowc$/$\medc$/$\highc$: low/medium/high dependency for Data, Prompt, and Ext.; unavailable/partial/full availability for Art.; for XGen., $\lowc$/$\highc$ indicate No/Yes. More filled squares in Repro. indicate easier reproduction.
  \item \textbf{Repro.}: more filled squares indicate easier reproduction, from difficult
  ($\blacksquare\square\square\square\square$) to deployment-ready
  ($\blacksquare\blacksquare\blacksquare\blacksquare\blacksquare$).
  \end{tablenotes}

  \end{threeparttable}
\end{table*}
\section{Defenses for LLM-Generated Phishing}\label{RQ3}
We examine defenses from two complementary perspectives: a systematization of existing defenses and an empirical evaluation of detector robustness across different LLM manipulation methods.

\subsection{Systematization Methodology}
% \noindent To assess how existing defenses address LLM-generated phishing, we map each defense to the nine generation stages and examine coverage across different manipulation methods (Table~\ref{tab:detection_methods_differ_rq3}). 

In Table~\ref{tab:detection_methods_differ_rq3}, we characterize existing defenses along five \textbf{Defense Objects:} a) Textual Characteristics Screening uses lexical, stylistic, and related textual features. b) Social Engineering Modeling captures persuasion, impersonation, urgency, and related social-engineering evidence. c) Intent Screening identifies malicious intent in prompts or generated content. d) Rule-Compliance Screening tests prompts, interactions, or outputs against safety policies rules. e) Human-Centric Defense addresses user susceptibility through training, awareness, and behavioral modeling.
% \textbf{Defense Scope:} 
% To characterize the \textbf{Defense Scope}, we record the \textbf{Defense Target}, indicating the aspect addressed by a defense: phishing content~\faEnvelope, content authorship~\faPenNib, malicious intent~\faBullseye, or user susceptibility~\faUser. We further record the \textbf{Defense Placement (Plc.)}, distinguishing prompt-level (\textbf{P}), post-generation (\textbf{O}), and user-side (\textbf{U}) defenses, which determines where intervention occurs relative to LLM-based phishing generation. Finally, \textbf{Stage Coverage (Stg.)} connects each defense to the RQ1 taxonomy by identifying the manipulation stages it addresses; coverage of one stage does not imply coverage of others.

We characterize \textbf{defense scope} through a) Defense Goal identifies the aspect addressed by a defense: phishing content~\faEnvelope, content authorship~\faPenNib, malicious intent~\faBullseye, or user susceptibility~\faUser. b) Defense Placement (Plc.) indicates where a defense operates relative to LLM-based phishing generation: prompt-level (\textbf{P}), post-generation (\textbf{O}), or user-side (\textbf{U}). c) Stage Coverage (Stg.) identifies the (\S\ref{RQ1}) manipulation stages that a defense can address based on the defense mechanism and evaluated attacks. Coverage of one stage does not automatically imply coverage of any other stage.

%\textbf{Defense Methodology:} To characterize defense designs, we record Defense Approach (App.) and Defense Model record used by each method, together with the Defense Evidence (Evd.) it relies on for detection or intervention. FL denotes federated learning and Cog. denotes cognitive or behavioral models.
\textbf{Defense Methodology:} To characterize defense designs, we record the Defense Approach (App.), the Defense Model used by each method, and the Defense Evidence (Evd.) on which it relies for detection or intervention. FL denotes federated learning, and Cog. denotes cognitive or behavioral models. 
% \textbf{c) Defense Evidence (Evd.)} records the information used by a defense for detection or intervention, including textual/stylometric evidence (ST), social-engineering cues (SE), prompt/request evidence (PR), external knowledge (KB), trigger/tag evidence (TAG), security policy (POL), and user behavioral evidence (BH). \textbf{Input Requirements:} We record the inputs required by each defense, including email headers (H), phishing communication text (C), and URL strings (U). Communication text includes email, SMS bodies, and Vishing scripts or messages.
\textbf{Operational Properties:} We further characterize existing defenses through training Data Dependency (Data) and Prompt Dependency (Prompt); evaluation across phishing generated by multiple LLMs (XGen.); the availability of code, datasets, models, and reproduction materials (Art.); reliance on external services, configurations, or infrastructure (Ext.); and overall Reproducibility (Repro.).
%\textbf{Operational Properties:} We further characterize existing defenses through training Data Dependency (Data) and prompt dependency (Prompt); whether defense is evaluated on phishing generated by multiple LLMs (XGen.); the availability of implementation resources such as the code, datasets, models, and reproduction materials (Art.); reliance on external services, configurations, or infrastructure (Ext.), and overall reproducibility (Repro.).
% \textbf{Operational Properties:} We assess the conditions required to apply, evaluate, and reproduce each defense along six dimensions. \textbf{a) Training Data Dependency (Data)} and \textbf{b) Prompt Dependency (Prompt)} reflect reliance on training data and prompt engineering, respectively ($\lowc$: low, $\medc$: medium, $\highc$: high).\textbf{c) Cross-Generator Evaluation (XGen.)} indicates whether a defense is evaluated on phishing generated by multiple LLMs ($\lowc$: No, $\highc$: Yes). \textbf{d) Artifact Availability (Art.)} records whether the code, datasets, models, and reproduction materials are publicly available  ($\lowc$: unavailable, $\medc$: partially available, $\highc$: fully available). \textbf{e) External Dependency (Ext.)} captures reliance on external services, configurations, or infrastructure ($\lowc$: largely standalone, $\medc$: moderate, $\highc$: strong). \textbf{f) Reproducibility (Repro.)} considers code, data, implementation details, and reproduction instructions, with more filled squares indicating easier reproduction.

\subsection{Existing Works Review}
\textbf{Content-centric defenses} analyze LLM-generated phishing using evidence ranging from textual patterns and social-engineering cues to generation intent and security policies.

\noindent \textbf{a) Textual Characteristics Screening} identifies LLM-generated phishing content from lexical and stylistic patterns in generated text. For instance, LLM-generated phishing exhibits lower punctuation frequency~\cite{greco2024david} and fewer spelling and grammar errors~\cite{gryka2024detection}, which are captured via diverse text representations, including TF-IDF~\cite{elongha2024detecting,brissett2025machine}, Word2Vec~\cite{wani2024ai}, Universal Data Analysis (UDAT)~\cite{eze2024analysis}, quantum-inspired feature encoding~\cite{malik2026qfedphish}, and LLM-reasoned phishing indicators~\cite{nahmias2024prompted}. These defenses are well studied for post-generation screening, where defenders analyze the phishing content without requiring access to the underlying LLMs.

\noindent \textbf{b) Social Engineering Modeling} identifies phishing by examining phishing evidence such as persuasion, impersonation, and urgency~\cite{bethany2025lateral,xue2025multiphishguard,nahmias2024prompted,heiding2024devising}. These methods are useful for context-rich phishing, where textual characteristics may provide insufficient detection evidence, while the expression retains cues of social-engineering strategies used to induce victim actions. For instance, Nahmias et al.~\cite{nahmias2024prompted} train the classifier only on traditional phishing and benign emails, which show substantial stylometric differences from LLM-generated phishing; however, the inferred social-engineering cues remain transferable to unseen phishing content.

% However, their reliability depends on whether the predefined semantic rules or prompts cover the relevant manipulation strategies. When LLM-generated phishing embeds intent in benign-looking contexts or uses softer requests, semantic cues alone may be insufficient for reliable discrimination.

\noindent \textbf{c) Intent Screening} leverages an additional defense point introduced by LLMs: the given prompts before content generation. Screening prompts enable phishing intent to be identified early, before the resulting content reaches downstream users and systems~\cite{roy2024chatbots}. Vafa et al.~\cite{vafa2026context} further enable detection of phishing-intent on the fabricated turns (Stg. = \textit{S5}). Pang et al.~\cite{pang2025paladin} use trigger-tag pairing to link prompt-side phishing triggers with output-side hidden tags, enabling defenders to identify malicious prompts and block generated content before delivery (Plc. =~\textit{P/O}). The methods are suitable for defenses deployed within LLM generation services, where user prompts are available before content generation. 
% These defenses enable intervention from \textbf{\textit{S1}} to \textbf{\textit{S6}}, while heavily relying on the training datasets (Data = \highc) and implementation configurations (Ext. = \highc).
% Additionally, these defenses remain vulnerable to benign-looking or obfuscated prompts that disguise phishing goals, such as \textit{``Draft a notification about an individual's eligibility for a prestigious credit card.''} (\textit{S6, S8}).

\begin{tcolorbox}[
  colback=gray!5, 
  colframe=black!100, 
  boxrule=1pt,
  arc=0pt,
  left=0mm,            
  right=0mm,            
  top=0.5mm,            
  bottom=0.5mm          
]
\textbf{Insight}: When complete prompt or interaction traces are available, Intent Screening enables intervention before delivery and can identify malicious intent that emerges across multiple attacker--LLM turns. Its protection is therefore limited to generation services that can observe those traces; downstream content-only defenses cannot rely on this evidence.
%\textbf{Insight} Intent Screening enables intervention before delivery and can capture phishing intent that emerges across multiple turns. However, these methods require access to generation prompts or interaction traces, which are unavailable to downstream content-only defenses.
% Prompt-side screening only works when the defender controls or observes the generation service. Future defenses should evaluate complete generation sessions and combine prompt-side screening with output-side phishing analysis.
\end{tcolorbox}

% \begin{tcolorbox}[
%   colback=gray!5, 
%   colframe=black!100, 
%   boxrule=0.3pt,
%   arc=0pt,
%   left=1mm,            
%   right=1mm,            
%   top=0.5mm,            
%   bottom=0.5mm          
% ]
% \textbf{Insight 11}: Prompt-level screening can intercept misuse whose malicious intent is observable in submitted prompts, but becomes less reliable when attackers distribute intent across turns, disguise it through benign framing, or shift the malicious objective to output rewriting or automated workflows.
% %\textbf{Insight 11}: Prompt-level screening can intercept early-stage misuse (S1--S3), but becomes less reliable when attackers distribute intent across turns, disguise it through benign framing, or move the malicious objective to later rewriting and automation stages.
% \end{tcolorbox}

\noindent \textbf{d) Rule-Compliance Screening} checks whether content or interactions conform to predefined rules and organizational procedures. Existing methods either rely on metadata and interaction patterns~\cite{mahendru2024securenet,liu2025pimref}, such as sender identities, URL domains, and conformance to organizational procedures; or evaluate LLM-generated content against governance, risk management, and compliance records~\cite{quinn2024applying}. However, this method may be less effective when predefined rules are insufficient to represent malicious intent expressed through semantic or contextual cues.

% \begin{tcolorbox}[
%   colback=gray!5, 
%   colframe=black!100, 
%   boxrule=0.3pt,
%   arc=0pt,
%   left=1mm,            
%   right=1mm,            
%   top=0.5mm,            
%   bottom=0.5mm          
% ]
% \textbf{Insight 12}: LLM-based analyzers provide interpretable reasoning beyond black-box classifiers, but their explanations may still reflect biased or incomplete reasoning patterns, such as overemphasizing urgency while underweighting contextual legitimacy.
% \end{tcolorbox}

% \begin{tcolorbox}[
%   colback=gray!15, 
%   colframe=black!100, 
%   boxrule=0.3pt,
%   arc=0pt,
%   left=1mm,            
%   right=1mm,            
%   top=0.5mm,            
%   bottom=0.5mm          
% ]
% \textbf{Gap 2}: Existing LLM-based analyzers lack mechanisms to verify whether their explanations faithfully reflect their decision logic. This creates a reliability risk: a detector may provide plausible reasoning while relying on spurious correlations, leading to unpredictable failures under adversarial or distribution-shifted inputs.
% \end{tcolorbox}

\textbf{Human-Centric Defense --} protects users by improving phishing awareness and detection capabilities, while modeling susceptibility to deceptive content. Malloy et al.~\cite{malloy2025training} model individual susceptibility (e.g., decisions, confidence, and actions) to predict how users react to LLM-generated phishing over time. These predictions can inform personalized training tailored to different user vulnerabilities. Human-centric defenses therefore facilitate individualized protection (Data = \highc, Repro. = $\blacksquare\square\square\square\square$), especially for user groups with well-characterized susceptibility patterns.

\begin{tcolorbox}[
  colback=gray!5, 
  colframe=black!100, 
  boxrule=1pt,
  arc=0pt,
  left=0mm,            
  right=0mm,            
  top=0.5mm,            
  bottom=0.5mm          
]
\textbf{Gap}: Of the 18 defenses summarized in Table~\ref{tab:detection_methods_differ_rq3}, 15 include post-generation placement, whereas only three include prompt-level placement and one explicitly models user susceptibility. This leaves generation-time misuse and user-specific risk comparatively undercovered.
%\textbf{Gap}: The defense landscape remains concentrated on post-generation textual proxies. Prompt- and interaction-level intent screening and defenses that incorporate user susceptibility are less represented, leaving limited coverage of generation-time misuse and user-specific risk.
% \tina{Fengchao: Please support ``concentrated'' and ``less represented'' with counts from Table~\ref{tab:detection_methods_differ_rq3}, such as the number of defenses in each Defense Object or Placement category. If the table does not support this comparison, please soften the wording rather than presenting it as a quantitative imbalance.}
%\textbf{Gap} Existing defense design primarily focuses on textual proxies of LLM-generated phishing, while phishing intent and user susceptibility receive substantially less attention.
\end{tcolorbox}

\begin{table*}[!t]
\caption{Detector MCC overview across LLM stages.}
\label{tab:mcc_overview_all_detectors}
\centering
\scriptsize
\setlength{\tabcolsep}{1pt}
\renewcommand{\arraystretch}{0.75}
\resizebox{\linewidth}{!}{%
\begin{tabular}{llccccccccccccccc}
\toprule
\multirow{2}{*}{Family}
& \multirow{2}{*}{Detector}
& \multirow[c]{2}{*}{\makecell[c]{HW\\MCC}}
& \multirow[c]{2}{*}{\makecell[c]{LLM\\MCC}}
& \multicolumn{13}{c}{Stage-aligned MCC on LLM-generated content} \\
\cmidrule(lr){5-17}
&
&
&
&
S1
& S2
& S3
& S4
& S6-MPG
& S6-UTA
& S6-Fuzzer
& S8-DeepSeek
& S8-GPT5.4
& S8-Gemini
& S8-Claude
& S8-Llama
& S8-Ministral \\
\midrule

\multirow{5}{*}{Academic}
& XGBoost
& 0.3467
& 0.2546
& 0.3461
& -0.0171
& 0.3081
& 0.7659
& 0.5794
& 0.1653
& 0.0049
& 0.1973
& 0.2311
& 0.1759
& 0.1485
& 0.2302
& 0.2253 \\

& T5-Phishing
& 0.0815
& -0.0444
& 0.0523
& 0.0200
& -0.0757
& -0.0307
& -0.0509
& -0.0713
& -0.1050
& -0.0917
& 0.4080
& 0.3316
& 0.2367
& -0.1324
& -0.1528 \\

& PiMRef
& 0.0831
& 0.0759
& 0.0209
& 0.0973
& 0.0304
& 0.0285
& 0.1028
& 0.0869
& 0.1017
& 0.0755
& 0.1553
& -0.0195
& 0.0996
& 0.1294
& 0.1513 \\

& ScamLLM
& 0.6324
& 0.4169
& 0.3857
& 0.1679
& 0.3946
& 0.8569
& 0.6967
& 0.5700
& 0.0122
& 0.3724
& 0.5261
& 0.5095
& 0.4736
& 0.5528
& 0.5308 \\

& SecureNet
& 0.7791
& 0.5276
& 0.3965
& 0.2607
& 0.6139
& 0.7710
& 0.8910
& 0.7356
& 0.4557
& 0.1962
& 0.3450
& 0.2502
& 0.1904
& 0.6416
& 0.6409 \\

\midrule
\multirow{5}{*}{Industrial}
& Email Agent
& 0.2174
& 0.2749
& 0.3282
& -0.0049
& 0.4246
& 0.8438
& 0.2120
& 0.2396
& 0.1573
& 0.3061
& 0.3233
& 0.3888
& 0.3428
& 0.3335
& 0.3176 \\

& Rspamd
& 0.2140
& 0.1932
& 0.3270
& -0.0413
& 0.4247
& 0.6618
& 0.3241
& 0.0943
& 0.1252
& 0.3213
& 0.0000
& 0.0000
& 0.0000
& 0.2460
& 0.2642 \\

& SpamScanner
& 0.1411
& 0.0972
& 0.0522
& -0.1090
& 0.0474
& 0.0404
& 0.1462
& 0.0777
& 0.1145
& 0.1320
& 0.0183
& 0.0257
& 0.0907
& 0.1442
& 0.0823 \\

& SpamAssassin
& 0.4514
& 0.3099
& 0.2591
& 0.0654
& 0.2624
& 0.5202
& 0.4309
& 0.2257
& 0.4180
& 0.2189
& 0.2151
& 0.1497
& 0.3611
& 0.1048
& 0.3169 \\

& PhishingV3
& 0.6331
& 0.4399
& 0.3619
& 0.0464
& 0.5496
& 0.8703
& 0.7398
& 0.6560
& 0.4545
& 0.0484
& 0.4969
& 0.4739
& 0.4986
& 0.4742
& 0.4855 \\

\bottomrule
\end{tabular}%
}
\end{table*}

\subsection{Benchmarking Phishing Detectors}\label{rq3_benchmark}
We next evaluate whether existing detectors remain robust as LLM-generated phishing changes across manipulation methods. We first establish overall performance differences, then examine detector disagreement and two controlled settings in which only the rewriting procedure or LLM generator changes.

\subsubsection{Benchmarking Setup}
\textbf{Detectors.} For reproducibility, we select detectors with publicly documented models, APIs, or executable deployment settings. Our benchmark includes five academic detectors: XGBoost~\cite{opara2025evaluating}, T5-Phishing~\cite{bethany2025lateral}, PiMRef~\cite{liu2025pimref}, ScamLLM~\cite{roy2024chatbots}, and SecureNet~\cite{mahendru2024securenet}; and five industrial detectors: Phishing Email Agent~\cite{phishing_email_agent}, Rspamd~\cite{Rspamd}, SpamScanner~\cite{spam_scanner}, SpamAssassin~\cite{spamassian_detector}, and PhishingV3~\cite{PhishingV3}. We evaluate the released detectors under their documented inference or deployment settings, without additional training, fine-tuning, or adaptation on the benchmark data.
%\textbf{Detectors.} For reproducibility, we select detectors with publicly documented models, APIs, or executable deployment settings. Our benchmark includes five academic detectors: XGBoost-phishing~\cite{opara2025evaluating}, T5-Phishing~\cite{bethany2025lateral}, PiMRef~\cite{liu2025pimref}, ScamLLM~\cite{roy2024chatbots}, and SecureNet~\cite{mahendru2024securenet}; and five industrial detectors: Phishing email agent~\cite{phishing_email_agent}, Rspamd~\cite{Rspamd}, SpamScanner~\cite{spam_scanner}, SpamAssassin~\cite{spamassian_detector}, and PhishingV3~\cite{PhishingV3}. We evaluate the released detectors under their documented inference or deployment settings, without additional training, fine-tuning, or adaptation on the benchmark data

\textbf{Datasets.} We use public datasets released within the past three years that contain phishing and/or benign content and are not reported as training or validation data for the evaluated detectors or generation methods. 
%\textbf{Datasets.} We use public phishing datasets from past three years that are not reported as training or validation data for the evaluated detectors or generation methods.
We map each sample to a benign or phishing label and distinguish human-written (HW) from LLM-generated sources. LLM-generated phishing subsets are assigned to the primary manipulation stage based on the reported generation procedure and metadata (\S\ref{RQ1}). We retain distinct settings within the same stage when they instantiate different generation procedures. For example, S6-MPG, S6-UTA, and S6-Fuzzer use different rewriting procedures, while the S8 settings distinguish LLM generators under the same workflow adapted from~\cite{roy2024chatbots}. Dataset composition and stage mapping are provided in Appendix~\ref{appendix:benchmarking_datasets}.
% \tina{Fengchao: Please re-audit every benchmark subset against the final stage definitions in Sec.~\ref{RQ1}. The appendix currently maps multi-turn data to S3, scenario data to S4, and targeted/BEC data to S5, which contradicts the final taxonomy: attacker--LLM multi-turn prompting is S5; scenario-, group-, or organization-level context is S3; and victim-specific or real-time target information is S4. Victim-facing multi-turn Vishing is S7. Please update the dataset metadata, Tables~\ref{tab:llm_phish_attr_overview}, \ref{tab:detection_methods_differ_rq3}, \ref{tab:mcc_overview_all_detectors}, and \ref{tab:stage_transfer_specificity}, all stage-wise prose, and the artifact together. Do not simply rename a column unless the underlying generation procedure supports the new label.}

\textbf{Metrics.} We report recall, precision, true negative rate (TNR), and MCC~\cite{matthews1975comparison} as the primary overall metrics. Recall measures phishing detection capability, TNR measures benign classification capability, and MCC reflects overall classification capability under class imbalance. 
% \tina{Fengchao: Please make the benchmark self-contained by reporting in the paper the exact detector versions, commits or API access dates, decision thresholds, per-subset counts after preprocessing by HW/LLM and benign/phishing label, deduplication and contamination checks, and any random seeds. The artifact can contain full manifests and commands, but it should not be the only place where reviewers can assess the evaluation protocol.}

\textbf{Notation.} Let $\mathcal{D}$ denote the benchmark samples. Each sample $x\in\mathcal{D}$ has a ground-truth label $y(x)\in{B,P}$ and source $\operatorname{src}(x)\in{\mathrm{HW},\mathrm{LLM}}$, where $B$ and $P$ denote benign and phishing content; $\operatorname{gen}(x)=g$ denotes its LLM generator when applicable. For detector $f$, $f(x)\in{B,P}$, yielding outcomes including true positive (TP) and false negative (FN). We use $\mathcal{D}_{s-y-o,g}$ to denote samples with source $s$, label $y$, outcome $o$, and generator $g$, omitting dimensions when not required (e.g., $\mathcal{D}_{\mathrm{LLM-P}}$ and $\mathcal{D}_{\mathrm{LLM-P-FN},g}$). For the SecureNet ($A$)–PhishingV3 ($I$) comparison, superscripts denote their joint outcomes, e.g., $\mathcal{D}_{\mathrm{LLM-P},g}^{\mathrm{A-TP,I-FN}}$. Finally, $r_p(x)$ denotes the persuasion score of pair $p$ for sample $x$.

\textbf{Analysis.} We annotate persuasion using the released model of Chen and Yang~\cite{chen2021weakly}. To capture more concrete expressions of phishing behavior, we consolidate recurring trigger words and phrases into linguistic and action-related characteristics using ChatGPT~\cite{steves2020categorizing,popovic2026dideco}: \textit{Urgency}, \textit{Login/Account}, \textit{Information Submission}, \textit{Click/Open}, \textit{Explicit Action}, \textit{Direct URL/Page Instruction}, \textit{Conversational Wording}, and \textit{Softened Requests}. This is followed by manual review and refinement.
 
Our analysis is backed by statistical tests. For detector outcomes on the same phishing inputs across LLM generators or rewriting methods, we use Cochran's $Q$ test~\cite{cochran1950comparison}, followed by McNemar tests~\cite{mcnemar1947note} when the overall difference is significant. For paired content comparisons, including different generations of the same source and matched HW--LLM samples, we use Friedman tests~\cite{friedman1937ranks} followed by Wilcoxon signed-rank tests~\cite{wilcoxon1945individual}. To compare content characteristics between independent detector-outcome groups (e.g., TP and FN), we use Mann--Whitney $U$ tests~\cite{mann1947test}. We apply Benjamini--Hochberg FDR~\cite{benjamini1995controlling} correction within each hypothesis family and report $q$-values.

% Because multiple characteristics are tested for each generator, we apply false discovery rate (FDR) correction to the $p$-values obtained from the Mann--Whitney $U$ tests (MWU, ~\cite{mann1947test}) and report the adjusted values as $q$-values. We present several representative statistically significant feature pairs in the paper, with the complete set of results available in our GitHub repository~\cite{sok_git}.

\subsubsection{Overall Benchmarking Results}  
Table~\ref{tab:mcc_overview_all_detectors} shows that nine of the ten evaluated detectors achieve lower MCC on the assembled LLM-generated subsets than on the HW subsets, although the magnitude of the difference varies substantially across detectors. For example, SecureNet, the highest-MCC detector on HW phishing, decreases from 0.7791 to 0.5276 on LLM-generated phishing, whereas Email Agent increases from 0.2174 to 0.2749.

\begin{figure}[ht]
    \centering
    \includegraphics[width=\linewidth]{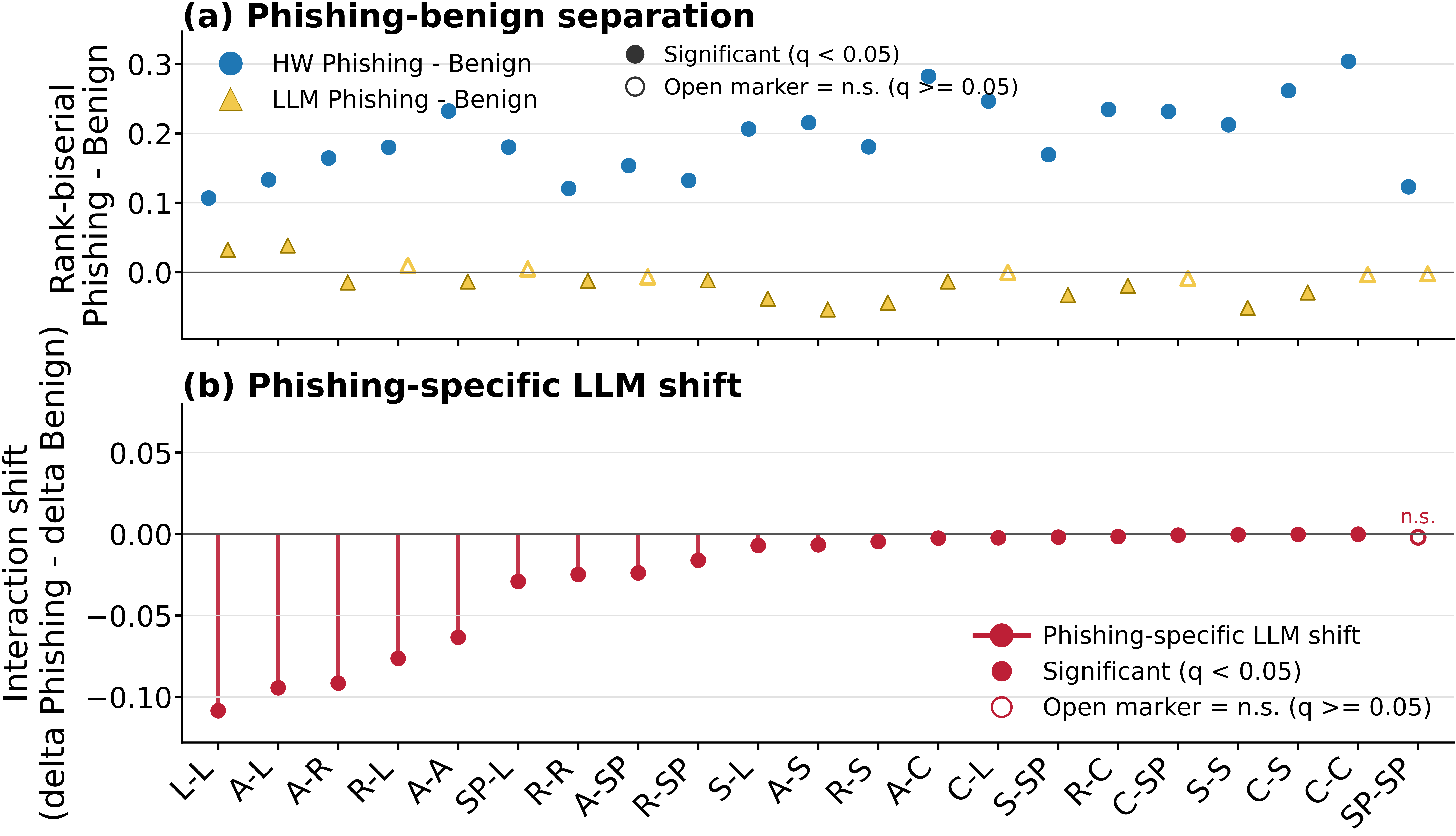}%,trim=0.7cm 0.3cm 0.7cm 0.3cm, clip
    \caption{Persuasion-pair separation across HW and LLM-generated content. (a) Rank-biserial effects for phishing versus benign content, reported separately for HW and LLM-generated samples. (b) Benign-controlled phishing-specific source shift $\kappa_p$. A, L, R, SP, S, and C denote Authority, Liking, Reciprocity, Social Proof, Scarcity, and Commitment, respectively (e.g., A-S = $(Authority, Scarcity)$). }
    \label{fig:e1_e2_split_compact}
\end{figure}

% Let $r_p(x)$ denote the score of persuasion pair $p$ in sample $x$. 
We first examine whether persuasion-pair characteristics provide similar separation between phishing and benign content across HW and LLM-generated text. 
For HW content, all reported persuasion pairs have significantly higher scores in phishing than in benign samples (all $q<0.001$, Fig.~\ref{fig:e1_e2_split_compact}a), with rank-biserial effects ranging from $0.107$ to $0.304$. This separation largely disappears for LLM-generated content. Only (\textit{Authority},\textit{Liking}) and (\textit{Liking},\textit{Liking}) retain significant positive differences, with small effects of 0.039 and 0.032. 
Twelve pairs instead show small but significant differences in the opposite direction, while seven show no significant difference (Fig.~\ref{fig:e1_e2_split_compact}a). 
%Twelve pairs instead show small but significant differences in the opposite direction, while seven show no significant difference (Fig.~\ref{fig:e1_e2_split_compact}b). 
These findings suggest that the discriminative patterns observed for HW phishing do not transfer uniformly to LLM-generated content.

We therefore further examine whether the HW-to-LLM shift in each persuasion pair differs between phishing and benign content, which allows us to identify changes that are more specific to LLM-generated phishing.
For each persuasion pair $p$, we define 
\begin{align} 
\delta_p^{P} &= \operatorname{median}_{x\in\mathcal{D}_{\mathrm{LLM-P}}} r_p(x) - \operatorname{median}_{x\in\mathcal{D}_{\mathrm{HW-P}}} r_p(x), \\ 
\delta_p^{B} &= \operatorname{median}_{x\in\mathcal{D}_{\mathrm{LLM-B}}} r_p(x) - \operatorname{median}_{x\in\mathcal{D}_{\mathrm{HW-B}}} r_p(x), 
\end{align} 
and measure the phishing-specific source shift as
\begin{equation} \kappa_p=\delta_p^{P}-\delta_p^{B}. \label{eq:phishing_specific_shift} \end{equation} 
A positive $\kappa_p$ indicates that the HW-to-LLM change is stronger in phishing content than in benign content, suggesting that this shift is more specific to LLM-generated phishing. We find that LLM generation increases persuasion-pair scores in phishing content ($q\approx0.001$), but the increase is not specific to phishing. For most persuasion pairs, the HW-to-LLM increase is even larger in benign content, such as (Liking, Liking) ($\delta_P=0.699$, $\delta_B=0.807$, $\kappa=-0.108$) and (Authority, Reciprocity) ($0.173$ vs.\ $0.265$, $\kappa=-0.092$). As a result, the persuasion-pair separation between phishing and benign content becomes smaller under LLM generation, making these features less discriminative for phishing detection.

We then examine whether the reduced phishing--benign separation is reflected in LLM-generated phishing being misclassified as benign. We compare persuasion-pair scores between false negatives and true positives for each detector, where a positive rank-biserial effect ($r_{\mathrm{rb}}$) indicates a higher pair score among phishing samples misclassified as benign. In general, the same persuasion-pair change is associated with different detection outcomes across detectors. For example, higher (\textit{Liking}, \textit{Liking}) scores are associated with more false negatives for SecureNet ($r_{\mathrm{rb}}=0.123$, $q<0.001$), but with more true positives for XGBoost ($r_{\mathrm{rb}}=-0.169$, $q<0.001$) and Rspamd ($r_{\mathrm{rb}}=-0.323$, $q<0.001$). For some detectors, LLM-generated phishing with increasingly benign-like persuasion patterns is more likely to be missed.

\begin{tcolorbox}[
  colback=gray!5, 
  colframe=black!100, 
  boxrule=1pt,
  arc=0pt,
  left=0mm,            
  right=0mm,            
  top=0.5mm,            
  bottom=0.5mm          
]
\textbf{Insight}: Across the assembled datasets, nine of ten detectors show lower MCC on LLM-generated than HW content, though this aggregate difference does not isolate LLM authorship from dataset shift. Persuasion-pair patterns that separate HW phishing from benign content also weaken or reverse for LLM-generated content, with detector-specific associations: more benign-like patterns yield more false negatives for some detectors but more true positives for others. Thus, HW performance does not establish robustness to LLM-generated phishing.
%\textbf{Insight}: Persuasion-pair patterns that separate human-written phishing from benign content provide weaker or reversed separation for LLM-generated content. Their associations with detector errors are detector-specific: more benign-like persuasion patterns correspond to more false negatives for some detectors but more true positives for others. Thus, persuasion-based separation established on human-written content does not by itself establish robustness to LLM-generated phishing.
%\textbf{Insight} The persuasion patterns that distinguish human-written phishing from benign content do not transfer directly to LLM-generated content. More importantly, LLM generation changes the persuasion expressions of both phishing and benign messages, narrowing the separation between the two classes; this reduced discriminability is reflected in the false-negative behavior of some detectors.
\end{tcolorbox}

\subsubsection{Representative Academic--Industrial Detector Disagreement}
Table 5 shows substantial variation and disagreement across detectors rather than a uniform performance gap between detector families.
%Table~\ref{tab:mcc_overview_all_detectors} shows lower MCCs for industrial detectors than the majority of academic detectors in our text-only benchmark. 
%To understand what content characteristics are associated with this detection gap,
To examine which content characteristics are associated with a representative cross-family disagreement, 
we compare two groups of phishing samples detected by SecureNet: those also detected by PhishingV3 ($\mathcal{D}_{\mathrm{LLM-P},g}^{\mathrm{A-TP,I-TP}}$) and those missed by PhishingV3 ($\mathcal{D}_{\mathrm{LLM-P},g}^{\mathrm{A-TP,I-FN}}$). SecureNet and PhishingV3 are the best-performing academic and industrial detectors in our benchmark, respectively. 
For each generator $g$ and characteristic $k$, we define $\Delta_{g,k}$ as the prevalence in $\mathcal{D}_{\mathrm{LLM-P},g}^{\mathrm{A-TP,I-FN}}$ minus that in $\mathcal{D}_{\mathrm{LLM-P},g}^{\mathrm{A-TP,I-TP}}$. $\Delta_{g,k}<0$ indicates higher prevalence among phishing detected by both detectors, whereas $\Delta_{g,k}>0$ indicates higher prevalence among phishing missed by PhishingV3.

% For each characteristic $k$, let $z_k(x)\in\{0,1\}$ indicate whether phishing instance $x$ contains characteristic $k$. We measure the difference as
% \begin{equation}
% \begin{aligned}
% \Delta_{g,k}
% ={}&
% \frac{1}{
% \left|\mathcal{D}_{\mathrm{LLM-P},g}^{\mathrm{A-TP,I-FN}}\right|}
% \sum_{x\in
% \mathcal{D}_{\mathrm{LLM-P},g}^{\mathrm{A-TP,I-FN}}}
% z_k(x)
% \\
% &-
% \frac{1}{
% \left|\mathcal{D}_{\mathrm{LLM-P},g}^{\mathrm{A-TP,I-TP}}\right|}
% \sum_{x\in
% \mathcal{D}_{\mathrm{LLM-P},g}^{\mathrm{A-TP,I-TP}}}
% z_k(x).
% \end{aligned}
% \label{eq:AI_feature_difference}
% \end{equation}
% Thus, $\Delta_{g,k}<0$ indicates higher prevalence among phishing detected by both detectors, whereas $\Delta_{g,k}>0$ indicates higher prevalence among phishing missed by PhishingV3.

\begin{figure}[!ht]
    \centering
    \includegraphics[width=\linewidth]{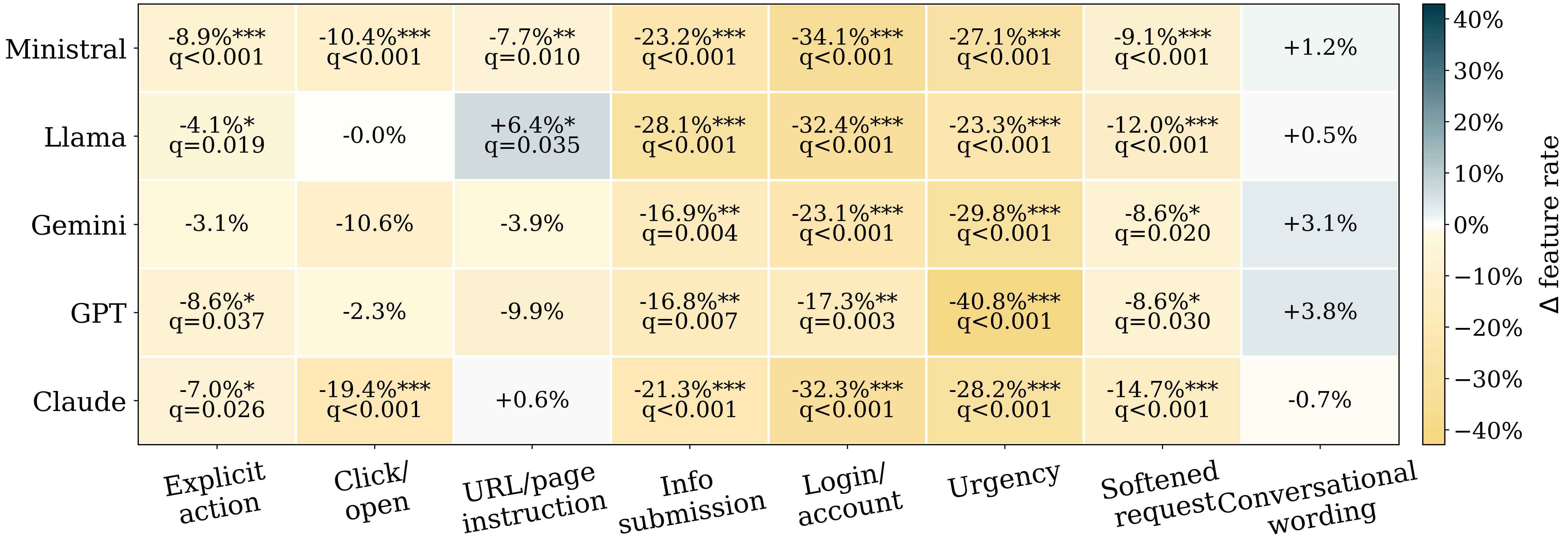}%,trim=0.7cm 0.3cm 0.7cm 0.3cm, clip
    \caption{Linguistic/action characteristic differences between \textit{A-TP, I-FN} and \textit{A-TP, I-TP} across LLM generators. Cells report $\Delta_{g,k}$, defined as \textit{A-TP, I-FN} prevalence minus \textit{A-TP, I-TP} prevalence, in percentage points (pp). DeepSeek is excluded because of a highly imbalanced comparison between the two groups.}
    \label{fig:AI_differences}
\end{figure}

Fig.~\ref{fig:AI_differences} shows the consistent differences in security-sensitive action cues. Across all five analyzed generators, urgency is $23.3\sim40.8$~pp less prevalent among phishing missed by PhishingV3 than among phishing detected by both detectors (all $q<0.001$). The same pattern holds for login/account cues ($\Delta_{g,k}\in[-34.1,-17.3]$~pp, $q\leq0.003$) and information-submission cues ($\Delta_{g,k}\in[-28.1,-16.8]$~pp, $q\leq0.007$). In contrast, characteristics describing general writing style do not show the same pattern. Conversational wording differs by only $-0.7$--$3.8$~pp and is not significant for any generator. URL/page instructions are also inconsistent across most generators, with no significant difference. These results indicate that the observed detector disagreement is more clearly reflected in explicit action-related characteristics than in general conversational style.
% urgency is $23.3\sim40.8$ percentage points (pp) less prevalent in $\mathcal{D}_{\mathrm{LLM-P},g}^{\mathrm{A-TP,I-FN}}$ than in $\mathcal{D}_{\mathrm{LLM-P},g}^{\mathrm{A-TP,I-TP}}$ across all five analyzed generators (all $q<0.001$). Login/account cues show differences of $17.3\sim34.1$~pp ($q\leq0.003$), while information-submission cues differ by $16.8\sim28.1$~pp ($q\leq0.007$). Other characteristics show less consistent patterns across generators. Conversational wording differs by only $-0.7\sim3.8$~pp and shows no significant difference for any generator. URL/page instructions vary in both magnitude and direction, from $-7.7$~pp for Ministral to $+6.4$~pp for Llama, with no significant difference for most generators. Among the measured characteristics, SecureNet--PhishingV3 disagreement is therefore most consistently associated with the explicit expression of security-sensitive actions, rather than with conversational wording.

\begin{tcolorbox}[
  colback=gray!5, 
  colframe=black!100, 
  boxrule=1pt,
  arc=0pt,
  left=0.5mm,            
  right=0mm,            
  top=0mm,            
  bottom=0.5mm          
]
\textbf{Insight}: For the representative academic and industrial detectors examined in our feature-level comparison, detection disagreement is more consistently associated with explicit action-oriented characteristics, particularly urgency, login/account requests, and information submission, than with general conversational wording.
% \fc{limitation: It remains unclear  whether the same pattern persists when deployment-side information such as sender reputation, headers, infrastructure, or communication history is available.}
\end{tcolorbox}

\subsubsection{Stage-aligned Benchmarking Results} 
The stage-level MCC results in Table~\ref{tab:mcc_overview_all_detectors} show that detector robustness varies substantially across LLM-generation settings. We examine representative settings that produce the largest or most informative performance differences. 

\textit{\textbf{Phishing Rephrasing}}. 
We compare three S6 settings that rewrite phishing content in different ways: \textit{S6-Fuzzer} allows relatively free LLM rewriting while preserving entity consistency~\cite{afane2024next}, \textit{S6-UTA} mainly applies synonym-based lexical replacement~\cite{utaliyeva2023chatgpt}, and \textit{S6-MPG} follows predefined rewriting rules~\cite{xue2025multiphishguard}. 
Using the same 1,600 source phishing messages across Fuzzer, UTA, and MPG, we find significant differences in detector outcomes among the three rewriting strategies ($q<0.001$), showing that the performance variation is attributable to the rewriting procedure rather than differences in source content. Among rewriting methods, Fuzzer produces broader shifts in persuasion relationships than UTA and MPG, increasing multiple pairs such as $(Authority, Liking)$, $(Authority, Reciprocity)$, and $(Liking, Social Proof)$, while UTA and MPG generally produce smaller or negative shifts across many of these persuasion pairs.
% \begin{figure}[!ht]
%     \centering
%     \includegraphics[width=0.98\linewidth]{diagrams/fig_s6_rewriting_two_panel_main_a.png}
%     \caption{Persuasion-pair shifts induced by three S6 rewriting methods relative to the same original phishing messages.}
%     \label{fig:s6_a}
% \end{figure}
% The \textit{Scarcity}-related persuasion relationships particularly strengthened by Fuzzer also show pronounced differences between its detected and missed phishing samples.

\begin{figure}[!ht]
    \centering
    \includegraphics[width=0.98\linewidth]{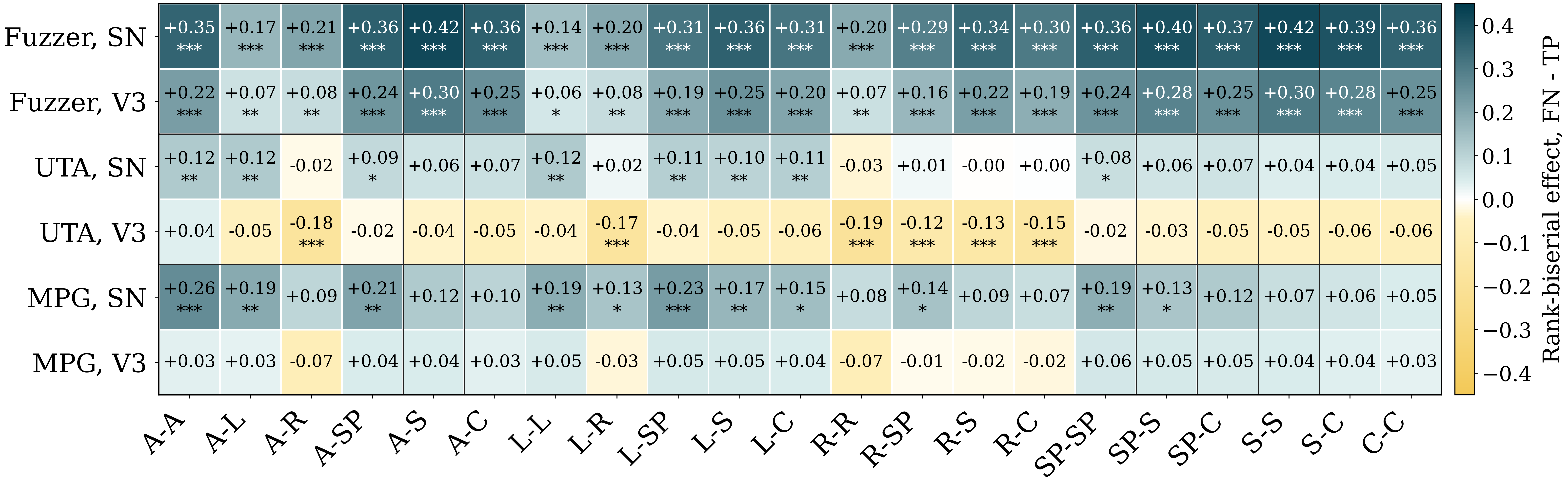}
    \caption{Persuasion-pair differences between detected and missed phishing under S6 rewriting. Cells report rank-biserial effect sizes ($r_{\mathrm{rb}}$), where positive values indicate higher scores among false negatives. A, L, R, SP, S, and C denote Authority, Liking, Reciprocity, Social Proof, Scarcity, and Commitment, respectively (e.g., A-S = $(Authority, Scarcity)$); SN and V3 denote SecureNet and PhishingV3.}
    \label{fig:s6_b}
\end{figure}

We then examine whether Fuzzer-strengthened persuasion pairs differ between detected and missed phishing and how these differences appear in concrete cues. We find that Fuzzer-induced changes in persuasion are reflected in detector failures, but the affected persuasion relationships differ across detectors. Fig.~\ref{fig:s6_b} shows that \textit{Scarcity}-related $(Authority, Scarcity)$ and $(Scarcity, Scarcity)$ pairs exhibit some of the strongest positive effects for SecureNet ($r_{\mathrm{rb}} \approx 0.42, q<0.001$). Notably, although \textit{Scarcity} is commonly expressed through urgency or time pressure, explicit urgency cues are $15.3$--$15.7$~pp less prevalent among false negatives than among true positives; login/account and information-submission cues are $35.8$--$38.2$~pp and $31.5$--$38.7$~pp less prevalent, respectively (all $q<0.001$; Fig.~\ref{fig:S6_combined}). In contrast, direct URL/page instructions show no significant difference for either detector, while click/open cues are inconsistent across detectors. Fuzzer-based rewriting then challenges detectors that rely on textual phishing cues by altering these cues while preserving phishing intent and requested actions.

% Finally, we examine how these Fuzzer failures manifest in concrete linguistic and action characteristics. Fig.~\ref{S6_combined} shows that $\mathcal{D}_{\mathrm{LLM-P-FN}}$ contains substantially fewer explicit phishing cues than $\mathcal{D}_{\mathrm{LLM-P-TP}}$ under both detectors. For instance, urgency is $15.3$--$15.7$~pp less prevalent in FN samples, and login/account cues are $26.9$--$38.2$~pp less prevalent (all $q<0.001$). In contrast, direct URL/page instructions show no significant difference for either detector, while click/open cues are inconsistent across detectors. These results show that Fuzzer failures combine stronger \textit{Scarcity}-related persuasion with fewer explicit phishing cues, indicating that the detector failures are associated with persuasive messages that retain strong \textit{Scarcity}-related structure while exposing fewer conventional action-oriented signals.

\begin{tcolorbox}[
  colback=gray!5, 
  colframe=black!100, 
  boxrule=1pt,
  arc=0pt,
  left=0.5mm,            
  right=0mm,            
  top=0mm,            
  bottom=0.5mm          
]
\textbf{Insight}: Holding the source phishing messages fixed, changing only the rewriting procedure significantly changes detector outcomes. For Fuzzer-based rewriting, detector failures are associated with stronger Scarcity-related persuasion and fewer explicit action cues, even though the underlying phishing objective is preserved. This suggests that detectors relying on surface textual cues may not generalize across rewriting strategies and motivates defenses that model the underlying phishing intent and requested action rather than their surface realization.
%\textbf{Insight}: Holding the source phishing messages fixed, changing only the rewriting procedure significantly changes detector outcomes. Under Fuzzer-based phishing rewriting, failures are associated with stronger Scarcity-related persuasion and fewer explicit action cues, even though the underlying phishing objective is preserved. This suggests that detectors relying on surface textual cues may not generalize across rewriting strategies and motivates evaluating intent- and action-oriented signals that remain consistent across rewritten variants. %This indicates that surface textual characteristics are not stable detection signals under rewriting; future defenses should instead identify the phishing intent and requested action that remain consistent across rewritten variants.
\end{tcolorbox}

\textit{\textbf{Diverse Generators}.}
%We first examine whether changing only the generator alters detector performance. We found that detection rates differ significantly across generators for all evaluated detectors (all $q<0.001$). However, the direction of this variation is detector-specific. 
Changing only the generator significantly changes detection rates for all evaluated detectors (all $q<0.001$), although the direction is detector-specific. 
For example, on $D_{LLM-P, Deepseek}$, SecureNet achieves a TPR of only $7.4\%$, while XGBoost achieves $87.8\%$ (Fig.~\ref{fig:S8_detection_rate}). %The results indicate that the detector performance is sensitive to generator choice, and robustness observed on a single LLM does not necessarily generalize to phishing generated by other LLMs.
Thus, robustness on one LLM does not necessarily generalize to phishing generated by others.
% \fc{Thus, changing only the generator significantly alters detector performance, but no generator is consistently the most difficult across detectors.}

\begin{figure}[ht]
    \centering
    \includegraphics[width=0.8\linewidth,trim=0cm 0cm 0cm 1cm, clip]{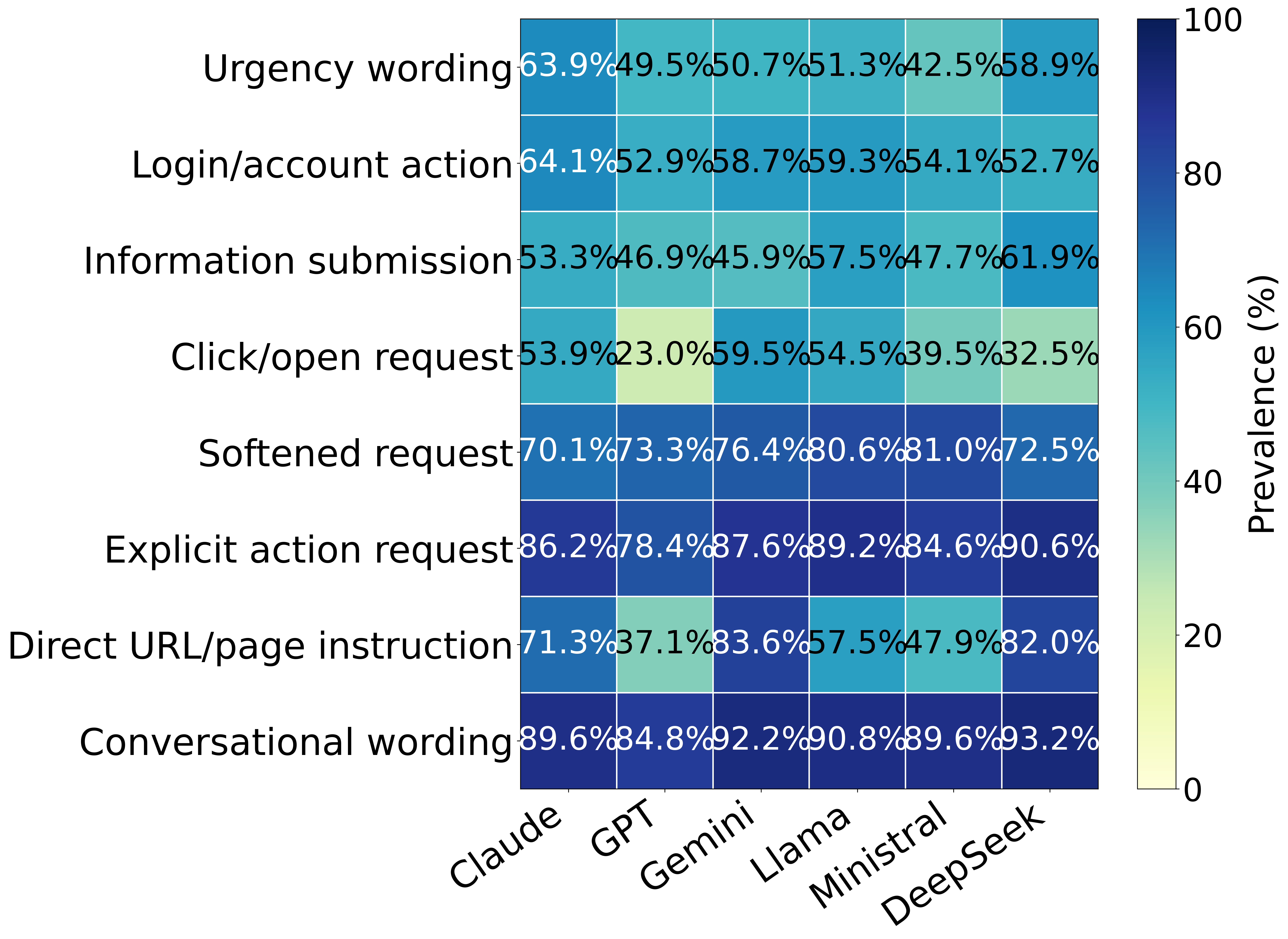}%
    \caption{Linguistic/action characteristic prevalence across LLM generators under S8. Cells report the percentage of generated phishing containing each characteristic.}
    \label{fig:s8_generator_features}
\end{figure}

% We then examine how the generated phishing content differs on linguistic and action-related characteristics across LLMs. The results show that all eight characteristics differ significantly across generators (all $q<0.001$), although the magnitude of the differences varies substantially. The largest variation occurs in direct URL/page instructions (Fig.~\ref{s8_generator_features}), ranging from $37.1\%$ for GPT to $83.6\%$ for Gemini ($46.5$~pp, $q<0.001$), followed by click/open requests from $23.0\%$ to $59.5\%$ for the same two generators ($36.5$~pp, $q<0.001$). 
% Thus, generator differences are concentrated particularly in how actionable phishing instructions are expressed rather than in general conversational wording.
% \begin{figure}[!ht]
%     \centering
%     \includegraphics[width=0.98\linewidth]{diagrams/Fig_S8_D_observed_vs_feature_adjusted_detection.png}
%     \caption{ .}
%     \label{fig:S6_combined}
% \end{figure}

% Figure~\ref{fig:s8_generator_features} summarizes the detector-specific differences in characteristic prevalence between true positives and false negatives.
We then examine how generator-dependent characteristics differ across LLMs. All eight characteristics differ significantly across generators (all $q<0.001$, Fig.~\ref{fig:s8_generator_features}), with the largest variations in action-related characteristics. Direct URL/page instructions range from 37.1\% for GPT to 83.6\% for Gemini (46.5 pp), while click/open requests range from 23.0\% to 59.5\% (36.5 pp; both $q<0.001$, Fig.~\ref{fig:s8_generator_features}). Together with the significant differences in detector outcomes across generators, these results show that generator choice changes both how phishing actions are expressed and how the resulting content is detected.

\begin{tcolorbox}[
  colback=gray!5, 
  colframe=black!100, 
  boxrule=1pt,
  arc=0pt,
  left=0.5mm,            
  right=0mm,            
  top=0mm,            
  bottom=0.5mm          
]
\textbf{Insight}: 
Holding prompts and generation procedure fixed, changing the LLM generator produces detector-specific performance changes. 
%Holding the prompts and generation procedure fixed, changing only the LLM generator significantly changes detector outcomes, and the direction of the change is detector-specific. 
Differences concentrate in how actionable phishing instructions are expressed, and several characteristics remain associated with detection after accounting for generator differences. 
%Generator-dependent differences are concentrated particularly in how actionable phishing instructions are expressed, and several of these characteristics remain associated with detection outcomes after accounting for generator differences. 
Performance on one generator therefore does not establish robustness to LLM-generated phishing generally.
% Changing the generative model can significantly alter detection performance, but it won't cause all phishing samples to change in the same direction overall. Different generators change particular properties of the phishing content, and different detectors respond differently to those properties.
\end{tcolorbox}

\section{Limitations and Research Directions}
Our corpus may be affected by publication and indexing bias, evolving preprints and model versions, and uneven reporting across phishing modalities. 
The nine stages form a non-exhaustive coding framework; hybrid methods may span stages, and assignments depend on the factor manipulated and evaluated in each study. 
%The nine stages are an evidence-based, non-exhaustive coding framework; hybrid methods may span stages, and primary-stage assignments depend on the manipulated and evaluated factor reported by each study. 
Our benchmark covers selected public datasets and text-focused detectors rather than deployment-wide effectiveness; undisclosed training data, dataset shift, label noise, and unavailable production signals may affect performance, while limited Smishing, Quishing, and Vishing data constrain cross-modal conclusions. 
%Our benchmark evaluates selected public datasets and text-focused detector configurations rather than deployment-wide effectiveness. Undisclosed training data, dataset shift, label noise, and unavailable production signals may affect observed performance, while limited Smishing, Quishing, and Vishing data constrain cross-modal conclusions. 
Feature analyses establish associations, not causality, motivating stage-aware benchmarks with transparent provenance, longitudinal evaluation, and dedicated multi-turn and non-email datasets.
%Feature analyses therefore establish associations, not causality. These limitations motivate stage-aware benchmarks with transparent provenance, longitudinal evaluation, and dedicated multi-turn and non-email textual phishing datasets.

\section{Conclusion}\label{conclusion}
%In this work, we have presented the first systematization of knowledge on LLM-generated phishing, covering the content-based phishing attack lifecycle from generation to mitigation. We believe our work represents the first comprehensive survey examining LLM-enabled phishing in an end-to-end manner. The work provides a structured overview of the literature, mapping categorizations of characteristics and defending methods aligning with building blocks, generation mechanisms. Our results indicate a change in the attack objectives of LLM-authored phishing. We also report an outdated development of defense methods compared with offenses, raising concerns about constructing resilient, adaptive, and robust defenses against systems. Finally, our analysis also suggests key insights and identifies research gaps, addressing existing constraints and guiding future directions. 

%In this work, we systematized LLM-generated phishing across content generation, reported characteristics and user-side effects, and content-focused defenses. 
We systematized LLM-generated phishing across generation methods, reported characteristics and user-side effects, and content-focused defenses. Our nine-stage framework organizes methods by the primary LLM-related factor under attacker control without imposing a sequence or capability ranking.
%Our nine-stage framework organizes manipulation methods by the primary LLM-related factor under attacker control without treating the stages as a required sequence or capability ranking. 
%We further mapped defense coverage to these stages and benchmarked five academic and five industrial detectors across heterogeneous LLM-generated phishing settings.
We mapped defense coverage to these stages and benchmarked five academic and five industrial detectors across heterogeneous settings. 
%The results show that detector performance varies across generators and rewriting strategies; several action-oriented characteristics are associated with detector outcomes, while persuasion-pair associations differ across detectors. 
Detector performance varies across generators and rewriting strategies; action-oriented characteristics are associated with detector outcomes, while persuasion-pair associations differ across detectors. 
%These findings are correlational and specific to the evaluated datasets and text-focused configurations; they do not imply that all LLM-generated phishing is harder to detect or that deployed defenses are uniformly static. 
%These correlational findings are specific to the evaluated datasets and text-focused configurations and do not imply that LLM-generated phishing is uniformly harder to detect or that deployed defenses are uniformly static. Overall, evidence and defense coverage remain limited, particularly for multi-turn and cross-modal settings.
These correlational findings are specific to our benchmark and do not imply that LLM-generated phishing is uniformly harder to detect or that deployed defenses are uniformly static.

\section*{Ethics Considerations}
This work is a systematization of knowledge (SoK) on LLM-generated phishing. Our analysis relies on prior published research and, where datasets are used, on publicly available resources cited in the paper; we follow the applicable licensing conditions. We do not collect new phishing data from real users or release executable attack pipelines. The purpose of this work is defensive, aiming to clarify the threat landscape and support the design of more resilient detection and defense mechanisms. We believe the study complies with the ethics guidelines by minimizing risks of misuse.

\section*{LLM usage considerations}
%We used Large Language Models (LLMs) only for editorial assistance to improve grammar, phrasing, and clarity of author-written text. All ideas, analyses, and conclusions are our own, and all LLM outputs were carefully reviewed and verified by the authors for accuracy, originality, and proper citation of prior work.
We used Large Language Models (LLMs) for editorial assistance to improve grammar, phrasing, and clarity of author-written text. We also used ChatGPT in the limited analysis step described in \S\ref{rq3_benchmark} to consolidate recurring trigger words and phrases into higher-level linguistic and action-related characteristics. The authors manually reviewed and refined all resulting categories and are responsible for all analyses, conclusions, accuracy, originality, and citations.

\section*{Open Science}\label{appendix:open_science}
An anonymized artifact for this submission is available at
\url{https://anonymous.4open.science/r/ssc_LLM_phishing_survey-BF08/README.md}.
It contains the study corpus and coding metadata, stage assignments, benchmark dataset manifests and preprocessing materials, detector configurations, analysis code, and instructions for reproducing the reported tables and figures. Security-sensitive reproduced phishing samples for \textit{S6} and \textit{S8} are not publicly released because of their misuse potential; the artifact documents these omissions and the materials available for review.

% \tina{Fengchao: Open Science is mandatory for USENIX Security '27. Please verify that this anonymous, non-tracking URL contains no identifying information or identifying commit history and remains accessible through the shepherd-approval deadline. Make this paragraph match the artifact exactly. Please check the study list and coding metadata, stage-assignment files, dataset sources, acquisition dates, licenses and class counts, preprocessing, exact detector versions or commits and API access dates, prompts and LLM model/access dates, random seeds, dependencies, commands, expected outputs, and the materials needed to reproduce every reported table and figure. If the S6/S8 samples are omitted even from confidential review, state the adversarial-risk rationale explicitly. Remove this comment before submission.}

\bibliographystyle{IEEEtran}
\bibliography{sample-base}



% \end{thebibliography}

\appendix

%\section*{Appendix}
\renewcommand{\thetable}{A\arabic{table}} 

\section{Benchmarking Used Datasets}\label{appendix:benchmarking_datasets}
We use public datasets covering different LLM manipulation stages identified in (\S\ref{RQ1}). We assign a stage only when the documented generation procedure identifies the corresponding manipulated factor. S1 includes direct malicious generation requests, while S2 covers role- or framing-based prompting. S3 includes scenario-, group-, or organization-level contextualization, whereas S4 requires victim-specific or real-time target information. S5 covers attacker--LLM multi-turn interactions that decompose or progressively elicit phishing content. S6 includes rewriting or paraphrasing that modifies generated content while preserving phishing intent. S7 covers manipulation of victim-facing communication form or interaction structure, including Vishing or Quishing only when the communication structure itself is manipulated. S8 includes automated workflows that coordinate prompt generation, evaluation, refinement, or selection. For our S8 reconstruction based on~\cite{roy2024chatbots}, we hold the workflow fixed and vary only the target LLM; generator identity is therefore treated as an attack property rather than a separate stage.

\section{Supplementary Figures and Tables}\label{appendix:why_Inds_bad}
\begin{figure}[!ht]
    \centering
    \includegraphics[width=0.8\linewidth]{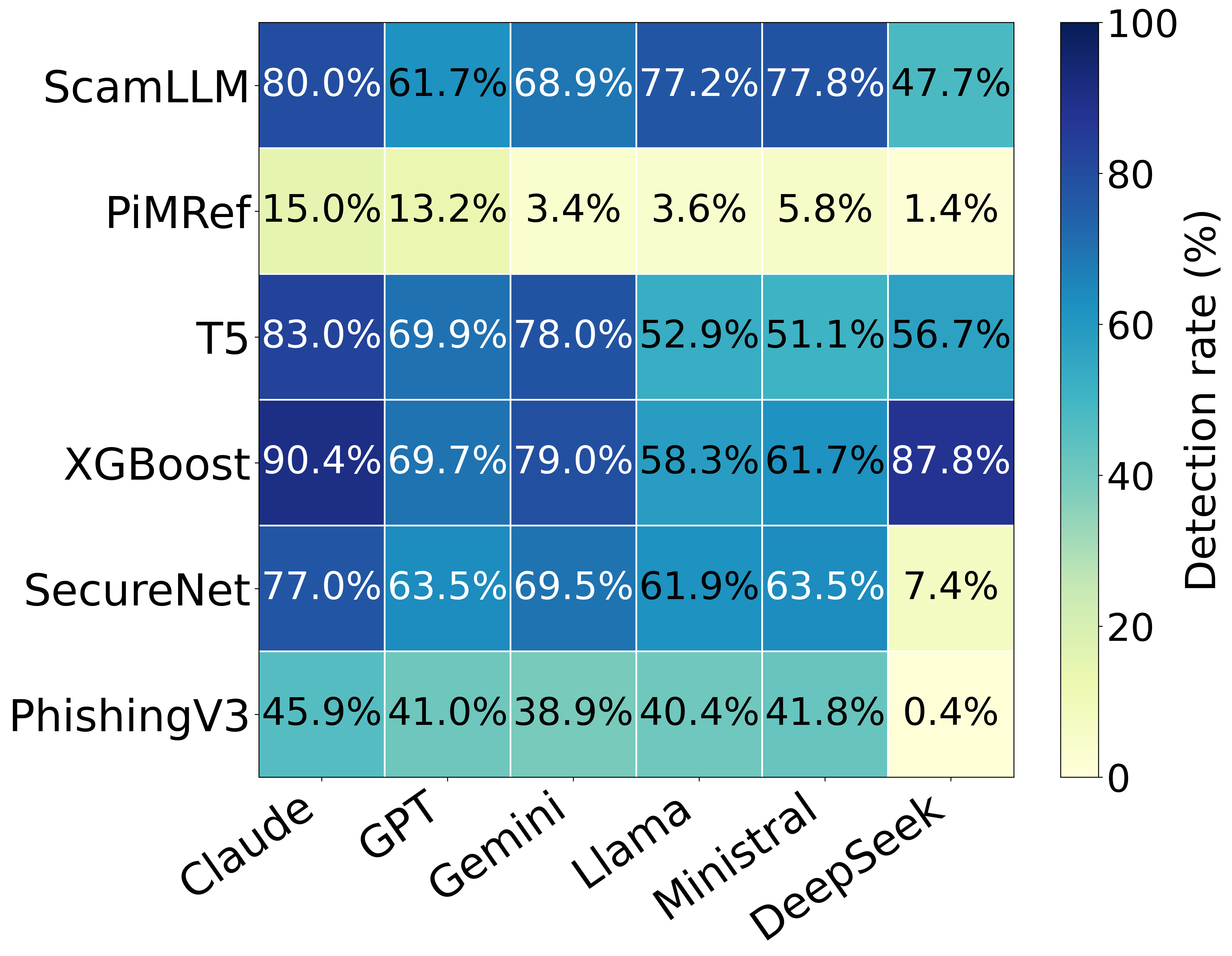}
    \caption{Detection rates across LLM generators under S8. Cells report the true-positive rate (\%) for each detector–generator combination.}
    \label{fig:S8_detection_rate}
\end{figure}

\begin{figure}[!ht]
    \centering
    \includegraphics[width=0.98\linewidth]{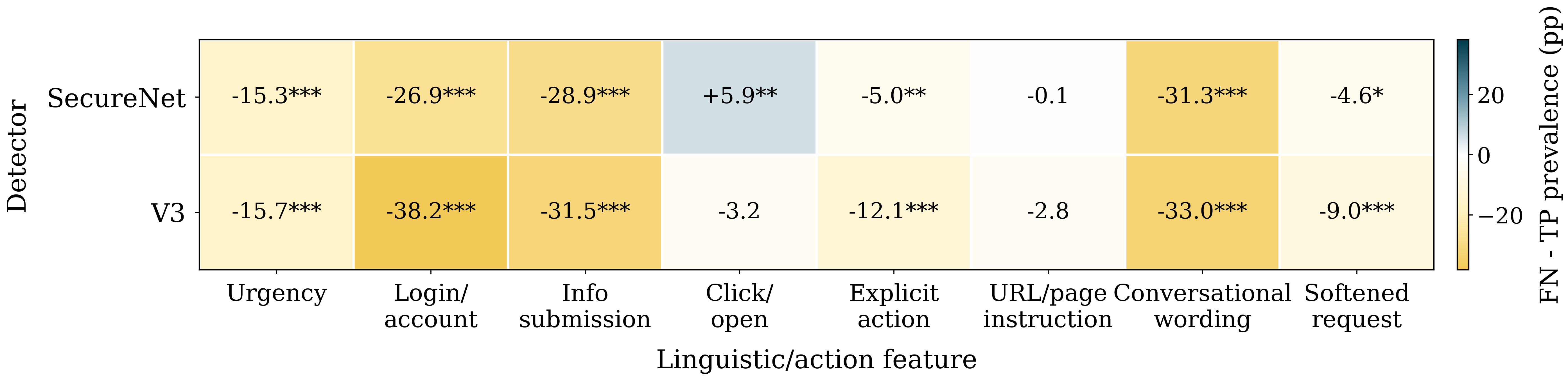}
    \caption{Linguistic/action characteristic differences between false negatives and true positives under S6-Fuzzer. Negative values indicate higher prevalence among correctly detected phishing.}
    \label{fig:S6_combined}
\end{figure}

% \begin{figure}[!ht]
%     \centering
%     \includegraphics[width=\linewidth]{diagrams/Fig_S8_C_detector_specific_tp_fn_feature_differences.png}
%     \caption{Detector-specific differences in characteristic prevalence between $D_{LLM-p-TP}$ and $D_{LLM-p-FN}$.}
%     \label{fig:S8_C}
% \end{figure}
\end{document}